\documentclass[twocolumn]{aastex6}

\usepackage{amssymb}
\usepackage{amsfonts}
\usepackage{amsmath}
\makeatletter
\def\env@matrix{\hskip -\arraycolsep 
  \let\@ifnextchar\new@ifnextchar
  \array{*{\c@MaxMatrixCols}c}}
\makeatother
\usepackage[varg]{txfonts}
\usepackage{natbib}
\usepackage{color}
\usepackage{graphicx}
\usepackage{hyperref}
\usepackage{tikz}
\usetikzlibrary{arrows.meta}
\usepackage{tikz-3dplot}
\usepackage[percent]{overpic}
\usepackage{multirow}

\let\oldint\int
\renewcommand{\int}[0]{\oldint\displaylimits}
\let\oldiint\iint
\renewcommand{\iint}[0]{\oldiint\displaylimits}
\newcommand{\prm}[1]{#1^\prime}
\newcommand{\pprm}[1]{#1^{\prime\prime}}

\newcommand{\vect}[1]{\boldsymbol{#1}}
\newcommand{\mrm}[1]{\mathrm{#1}}
\newcommand{\trm}[1]{\textrm{#1}}

\newcommand{\tpihlf}[0]{{\textstyle\frac{\pi}{2}}}
\newcommand{\tpieght}[0]{{\textstyle\frac{\pi}{8}}}

\newcommand{\AxisRotator}[1][rotate=0]{%
    \tikz [x=0.15cm,y=0.60cm,line width=.25ex,-stealth,#1,solid] \draw (0,0) arc (-150:150:1 and 0.5);%
}

\defcitealias{Almanstoetteretal18}{AMJM18}

\shorttitle{Generalized 3D gravity solver on a spherical multi-patch grid}

\shortauthors{A.~Wongwathanarat}

\begin{document}

\title{A Generalized Solution Method for Parallelized Computation of the 
Three-dimensional Gravitational Potential on a Multi-patch grid in Spherical Geometry}

\author{Annop Wongwathanarat}

\affil{Max-Planck-Institut f\"ur Astrophysik,
       Karl-Schwarzschild-Str. 1, 85748 Garching, Germany}
\email{annop@mpa-garching.mpg.de}

\begin{abstract}
We present a generalized algorithm based on a spherical harmonics expansion method
for efficient computation of the three-dimensional gravitational potential on a multi-patch 
grid in spherical geometry. Instead of solving for the gravitational potential by superposition 
of separate contributions from the mass density distribution on individual grid patch our new 
algorithm computes directly the gravitational potential due to contributions from all grid 
patches in one computation step, thereby reducing the computational cost of the gravity 
solver. This is possible by considering a set of angular weights which are derived from  
rotations of spherical harmonics functions defined in a global coordinate system that is 
common for all grid patches. Additionally, our algorithm minimizes data communication 
between parallel compute tasks by eliminating its proportionality to the number of subdomains 
in the grid configuration, making it suitable for parallelized computation on a multi-patch 
grid configuration with any number of subdomains. Test calculations of the gravitational 
potential of a tri-axial ellipsoidal body with constant mass density on the Yin-Yang 
two-patch overset grid demonstrate that our method delivers the same level of accuracy 
as a previous method developed for the Yin-Yang grid, while offering improved computation 
efficiency and parallel scaling behaviour.
\end{abstract}

\keywords{methods: numerical --- gravitation}

\section{Introduction}

When modeling self-gravitating systems a gravity solver that
calculates the gravitational potential by solving the Poisson's
equation is one of the central components in the simulations. For
multi-dimensional hydrodynamical simulations of self-gravitating flows
the Poisson's equation is solved at every hydrodynamical time step,
and thus can be responsible for a significant fraction of the
computational cost. Hence, several techniques for an efficient gravity
solver have been developed over the past decades. An algorithm of
choice is often decided by the complexity of the mass density
distribution on the computational domain and the numerical technique
used to solve the coupled hydrodynamic equations. For instance,
hierarchical tree-based algorithm
\citep[e.g.,][]{Appel85,Jernigan85,Porter85,BarnesHut86,HernquistKatz89}
is usually the preferred choice in particle-based codes due to its
flexibility in considering arbitrary geometry. On the other hand, for
grid-based codes a solver based on Fast Fourier Transform (FFT) can
easily be applied when the grid spacing is uniform \citep{Hockney70,
  BorisRoberts69}. For problems with large spatial dynamic ranges such
as cosmological structure formation or star cluster formation an
adaptive mesh refinement (AMR) technique is employed in order to
achieve high effective spatial resolution. On such adaptive grids
iterative multi-grid gravity solvers \citep[e.g.,][]{Ricker08} are
usually used, but an application of the tree-based solver on AMR grids
has also recently been investigated \citep{Wuenschetal18}. 

In stellar hydrodynamics, three-dimensional (3D) simulations have
mostly been performed on a spherical polar grid. For this class of
simulations a common choice of gravity solvers is the algorithm based
on a spherical harmonics expansion of the Green's function developed
by \citet{MuellerSteinmetz95}. This gravity solver has been employed,
for example, in several 3D CCSNe simulations
\citep[e.g.,][]{Vartanyanetal19,Glasetal18,Wongwathanaratetal17,Lentzetal15}
due to its high computation efficiency in a case where the
gravitational potential is dominated by the monopole term contribution
from a central quasi spherical body (e.g., the proto-neutron star in
CCSNe). It has also been adapted for a calculation on a Cartesian mesh
\citep{Couchetal13}, and is now implemented as a modular component in
recent versions of the {\sc Flash} code
\citep{Fryxelletal00,Dubeyetal09}. On the other hand, an alternative
algorithm which solves the discretized Poisson's equation on a
spherical polar grid using FFT has recently been developed by
\citet{MuellerChan18}. This algorithm gives more accurate solutions of
the gravitational potential than those obtained by the multipole
expansion technique for extremely asymmetric density configurations
such as an off-center point mass. Such an algorithm will particularly
be more advantageous in cases where multiple components of mass
concentration are present on the grid. 

Because spatial discretization of the computational domain by a
spherical polar grid introduces a severe time step constraint imposed
by small grid zones in the polar regions, modern multi-patch grid
techniques in spherical geometry such as the Yin-Yang grid
\citep{KageyamaSato04} and the cubed-sphere grid \citep{Ronchietal96}
have recently been receiving more attention. These grid techniques
avoid coordinate singularities at the poles, and therefore help to
speed up simulations by increasing the allowed time step size. At the
Garching supernova group, the Yin-Yang grid, which is a two-patch
overset grid configuration, is implemented into the finite-volume
neutrino-radiation hydrodynamic code {\sc Prometheus-Vertex}
\citep{Fryxelletal89,RamppJanka02}, and is now being used extensively
by  for performing state-of-the-art calculations of core-collapse
supernovae (CCSNe) in 3D \citep{Summaetal18,Melsonetal15}.

Computation of self-gravity on a multi-patch grid in spherical
geometry is non-trivial. Extensions of the spherical harmonics solver
by \citet{MuellerSteinmetz95} and the FFT-based solver by
\citet{MuellerChan18} for a multi-patch grid are not readily
available. For the case of the Yin-Yang grid,
\citet{Wongwathanaratetal10} resorted to a simple approach. They
interpolated the density field on the Yin-Yang grid onto an auxilliary
spherical polar grid, and then applied the algorithm by
\citet{MuellerSteinmetz95} without any modification. While this
approach provides an easy workaround to the solution, it is not ideal
since it introduces an additional source of errors through
interpolation of the density field. Moreover, the method cannot be
easily and efficiently paralellized for computation with a large
number of processes because of complicated data communication
pattern. 

A more efficiently parallelized approach for solving the gravitational
potential directly on the Yin-Yang grid based on the algorithm by
\citet{MuellerSteinmetz95} has recently been proposed by
\citet[][hereafter AMJM18]{Almanstoetteretal18}. In their method, the
gravitational potential is computed by adding contributions from the
mass density distribution on the Yin and the Yang grid section, which
are evaluated separately. Since each grid patch in the Yin-Yang grid
configuration is simply the low-latitude part of the usual spherical
polar grid{\bf,} solving for the gravitational potential using this
approach is straightforward. Although this method eliminates the need
for interpolation of the density field onto a spherical polar grid and
is easily parallelized on distributed-memory systems, it is still not
optimal. On the one hand, the computational cost is increased because
it calculates two sets of potential. On the other hand, the data
communication volume among parallel compute tasks is also enlarged by
a factor of two when compared to computation on a spherical polar grid
with the same number of compute tasks. The latter imposes a limit to
the parallel scalability of the algorithm. Furthermore and most
importantly, an extension of this algorithm for other multi-patch
configurations in spherical geometry with a larger number of grid
sections (e.g., the cubed-sphere grid) would significantly decrease
its computational efficiency because the additional computational cost
and the size of data communication are proportional to the total
number of grid patches.

In this paper, we derive a new algorithm for efficient computation of the 3D gravitational 
potential on a multi-patch grid in spherical geometry based on spherical harmonics expansion. 
Our method is a generalization of the method by \citet{MuellerSteinmetz95}. It differs from 
the previous algorithm by \citetalias{Almanstoetteretal18} in that our method calculates 
the sum of all contributions to the gravitational potential from all grid patches in one 
computation step. It takes full advantage of the symmetry property of the multi-patch 
grid configuration when calculating angular weights for the density distribution on 
each grid patch by utilizing rotational transformations of spherical harmonics. Data 
communication between compute tasks in parallel computation is minimized such that 
there is no dependency on the number of grid patches in the mesh
configuration. Consequently, this gravity solver is suitable not only
for the Yin-Yang grid with two grid sections but also for other
multi-patch configurations in spherical geometry that consist of a
larger number of grid patches. 

Our paper is organized as follows: We begin by summarizing the basic algorithm by 
\citet{MuellerSteinmetz95} for solving the 3D gravitational potential on a spherical 
polar grid in Section~\ref{sec:phi_spherical}. Then, we present our generalization of the 
basic algorithm for a multi-patch grid configuration in spherical geometry, and give 
explicit formulae of angular and radial weights needed for reconstruction of 
the gravitational potential in Section~\ref{sec:multi-patch-algo}. In 
Section~\ref{sec:YYimplementation}, we detail our implementation of the new algorithm for 
the case of computation on the Yin-Yang overset grid. Our algorithm is validated with a 
test computation on the Yin-Yang grid. The results are shown in Section~\ref{sec:results} 
along with a performance analysis of our algorithm in comparison with the previous 
method by \citetalias{Almanstoetteretal18}. We conclude with discussions in 
Section~\ref{sec:conclusions}.

\section{Base Algorithm on a spherical polar grid}
\label{sec:phi_spherical}

\subsection{Basic equations}
A brief summary of the efficient algorithm by \citet{MuellerSteinmetz95} for solving the integral 
form of the Poisson's equation is as follows. The Poisson's equation in its integral form
reads 
\begin{equation}
\Phi(\vect{r}) = -G\int\mrm{d}^3\vect{\prm{r}}
\frac{\rho(\vect{\prm{r}})}{\lvert\vect{r}-\vect{\prm{r}}\rvert}.
\label{eq:integralpoisson}
\end{equation}
Here, $G$ is the gravitational constant, $\vect{r}=(r,\theta,\phi)$ is a coordinate vector 
in spherical polar coordinates, and $\rho(\vect{r})$ is the density distribution function.
To solve this equation the Green's function $\lvert\vect{r}-\vect{\prm{r}}\rvert^{-1}$ is 
expanded into spherical harmonics. Following this expansion, the gravitational potential 
at a point $\vect{r}$ can then be expressed as 
\begin{equation}
\Phi(\vect{r}) = -G\sum_{\ell=0}^{\infty}\frac{4\pi}{2\ell+1}\sum_{m=-\ell}^{\ell}  
Y_\ell^m(\theta,\phi)
\cdot \left[\mathcal{A}_{\ell m}(r)+\mathcal{B}_{\ell m}(r)\right]
\label{eq:gravpotspherical}
\end{equation}
where the radius dependence functions $\mathcal{A}_{\ell m}$ and $\mathcal{B}_{\ell m}$ 
describing contributions to the gravitational potential from the mass distribution 
inside and outside of a radial coordinate $r$ are defined as 
\begin{equation}
\mathcal{A}_{\ell m}(r) = \frac{1}{r^{\ell+1}}\iint_{4\pi}\mrm{d}\prm{\Omega}Y_\ell^{m\,*}
(\prm{\theta},
\prm{\phi})\,\int_0^r\mrm{d}\prm{r}(\prm{r})^{\ell+2}\rho(\vect{\prm{r}})
\label{eq:A_lm}
\end{equation}
and
\begin{equation}
\mathcal{B}_{\ell m}(r) = r^\ell\iint_{4\pi}\mrm{d}\prm{\Omega}Y_\ell^{m\,*}
(\prm{\theta},
\prm{\phi})\,\int_r^\infty\mrm{d}\prm{r}(\prm{r})^{1-\ell}\rho(\vect{\prm{r}})
\label{eq:B_lm}
\end{equation}
with $\mrm{d}\Omega=\sin{\theta}\,\mrm{d}\theta\,\mrm{d}\phi$.
Here we use the definition 
\begin{equation}
Y_\ell^m(\theta,\phi) = \sqrt{\frac{2\ell+1}{4\pi}\frac{(\ell-m)!}{(\ell+m)!}}
\cdot P_\ell^m(\cos{\theta})\cdot e^{im\phi}
\label{eq:Y_l^m}
\end{equation}
for spherical harmonics of degree $\ell$ and order $m$ where $P_\ell^m$ are the associated 
Legendre polynomials. Complex conjugates of spherical harmonics $Y_\ell^m$ are denoted as 
$Y_\ell^{m\,*}$.

Using the definition in Eq.~(\ref{eq:Y_l^m}) and an identity for $P_\ell^m$ 
with negative orders,
\begin{equation}
P_\ell^{-m} = (-1)^m\frac{(\ell-m)!}{(\ell+m)!}P_\ell^m,
\label{eq:p_l^-m}
\end{equation}
Eq.~(\ref{eq:gravpotspherical}) can be re-written as
\begin{multline}
\Phi(\vect{r}) = -G\sum_{\ell=0}^{\infty}\sum_{m=0}^{\ell}\frac{2}{\delta_m}
\frac{(\ell-m)!}{(\ell+m)!}P_\ell^m(\cos{\theta}) \\
\cdot\left[\mathcal{G}_{\ell m}(r,\phi)+\mathcal{H}_{\ell m}(r,\phi)\right]
\label{eq:phispherical}
\end{multline}
where
\begin{multline}
\mathcal{G}_{\ell m}(r,\phi) = \frac{1}{r^{\ell+1}}\iint_{4\pi}\mrm{d}\prm{\Omega}
 P_\ell^m(\cos{\prm{\theta}}) \\
\times\cos{(m(\phi-\prm{\phi}))}\int_0^r\mrm{d}\prm{r}(\prm{r})^{\ell+2}\rho(\vect{\prm{r}})
\label{eq:C_lm}
\end{multline}
and 
\begin{multline}
\mathcal{H}_{\ell m}(r,\phi) = r^\ell\iint_{4\pi}\mrm{d}\prm{\Omega}
 P_\ell^m(\cos{\prm{\theta}}) \\
\times\cos{(m(\phi-\prm{\phi}))}\int_r^\infty\mrm{d}\prm{r}(\prm{r})^{1-\ell}\rho(\vect{\prm{r}}).
\label{eq:D_lm}
\end{multline}
The coefficient $\delta_m$ is defined by 
\begin{equation}
\delta_{m} = \begin{cases} 2, & \mrm{if}~m=0, \\
                          1, & \mrm{otherwise}. \end{cases}
\label{eq:del_m}
\end{equation}

\subsection{Discretized formulae on a spherical polar grid}
\label{sec:discretizedsph}

Let us construct a spherical polar grid of $N_r\times N_\theta\times N_\phi$ grid cells, 
and use indices $i$, $j$, and $k$ to label the $i^\mrm{th}$, $j^\mrm{th}$, and 
$k^\mrm{th}$ grid cell in the $r$-, $\theta$-, and $\phi$-direction, respectively. For each grid 
cell $\mathcal{Z}_{ijk}$ coordinates of the (lower)higher side of the cell in each coordinate 
direction are 
denoted by $r_i^{(-)+}$, $\theta_j^{(-)+}$, and $\phi_k^{(-)+}$. To compute the gravitational potential 
on this grid numerically using Eq.~(\ref{eq:phispherical}) we make two assumptions. Firstly, 
we assume that the density distribution inside a grid cell $\mathcal{Z}_{ijk}$ is constant, 
and is approximated 
by the cell-averaged density $\rho_{ijk}$. Secondly, we truncate the summation over spherical 
harmonic degree $\ell$ at a chosen value $\ell_\mrm{max}$.

Consider a case that the gravitational potential is to be calculated at grid zone 
interfaces in the radial direction, $\mathcal{G}_{\ell m}$ and $\mathcal{H}_{\ell m}$ 
can be numerically computed for the $n^\mrm{th}$ radial grid interface as 
\begin{equation}
  \mathcal{G}_{\ell m}(r_n^+,\phi) = \frac{1}{(r_n^+)^{\ell+1}}\sum_{i=1}^n\sum_{j=1}^{N_\theta}
\sum_{k=1}^{N_\phi}\rho_{ijk}\mathcal{R}_{\mrm{in},i}^{(\ell)}\mathcal{T}_j^{(\ell m)}\mathcal{F}_k^{(m)}
\label{eq:C_lm_sum}
\end{equation}
and
\begin{equation}
  \mathcal{H}_{\ell m}(r_n^+,\phi) = (r_n^+)^\ell\sum_{i=n+1}^{N_r}\sum_{j=1}^{N_\theta}\sum_{k=1}^{N_\phi}
\rho_{ijk}\mathcal{R}_{\mrm{out},i}^{(\ell)}\mathcal{T}_j^{(\ell m)}\mathcal{F}_k^{(m)}
\label{eq:D_lm_sum}
\end{equation}
where
\begin{equation}
  \mathcal{F}_k^{(m)} = \cos{(m\phi)}\,\mathcal{C}_k^{(m)}+\sin{(m\phi)}\,\mathcal{S}_k^{(m)}.
\label{eq:F_k^m}
\end{equation}
The angular weights $\mathcal{T}_j^{(\ell m)}, \mathcal{C}_k^{(m)},$ and 
$\mathcal{S}_k^{(m)}$ in Eqs.~(\ref{eq:C_lm_sum})--(\ref{eq:F_k^m}) are defined by  
\begin{equation}
\mathcal{T}_j^{(\ell m)} = \int_{\theta_j^{-}}^{\theta_j^{+}}\mrm{d}\prm{\theta}\sin{\prm{\theta}}
P_\ell^m(\cos{\prm{\theta}}),
\label{eq:T_lm}
\end{equation}
\begin{equation}
\mathcal{C}_k^{(m)} = \int_{\phi_k^{-}}^{\phi_k^{+}}\mrm{d}\prm{\phi}\cos{(m\prm{\phi})},
\label{eq:C_k^m}
\end{equation}
and
\begin{equation}
\mathcal{S}_k^{(m)} = \int_{\phi_k^{-}}^{\phi_k^{+}}\mrm{d}\prm{\phi}\sin{(m\prm{\phi})}.
\label{eq:S_k^m}
\end{equation}
The integrals $\mathcal{T}_j^{(\ell m)}$ can be computed analytically with help of recurrence 
formulae for the associated Legendre polynomials, while the integrals $\mathcal{C}_k^{(m)}$ and 
$\mathcal{S}_k^{(m)}$ are elementary. Analytic solutions to these integrals are already explicitly 
given by \citet{Zwerger95} and recently also in the work by \citetalias{Almanstoetteretal18}. 
Finally, the radial integrals 
\begin{align}
    \mathcal{R}_{\trm{in},i}^{(\ell)} = \int_{r_i^{-}}^{r_i^{+}}\mrm{d}\prm{r}(\prm{r})^{\ell+2}
\label{eq:R_in}
\end{align}
and
\begin{align}
    \mathcal{R}_{\trm{out},i}^{(\ell)} = \int_{r_i^{-}}^{r_i^{+}}\mrm{d}\prm{r}(\prm{r})^{1-\ell}
\label{eq:R_out}
\end{align}
are also easy to compute.

Following the implementation steps suggested by \citet{MuellerSteinmetz95} the angular 
parts of the summations in Eqs.~(\ref{eq:C_lm_sum}) and (\ref{eq:D_lm_sum}) defined by 
\begin{equation}
A_{C,i}^{(\ell m)}=\sum_{j=1}^{N_\theta}\sum_{k=1}^{N_\phi}\rho_{ijk}\mathcal{T}_j^{(\ell m)}\mathcal{C}_k^{(m)}
\label{eq:a_ci^lm}
\end{equation}
and
\begin{equation}
A_{S,i}^{(\ell m)}=\sum_{j=1}^{N_\theta}\sum_{k=1}^{N_\phi}\rho_{ijk}\mathcal{T}_j^{(\ell m)}\mathcal{S}_k^{(m)}.
\label{eq:a_si^lm}
\end{equation}
are computed first. Then, the radial summations can be evaluated efficiently by utilizing
recurrence relations
\begin{equation}
\sum_{\prm{i}=1}^{i}\mathcal{R}_{\mrm{in},\prm{i}}^{(\ell)}A_{C/S,\prm{i}}^{(\ell m)} =
\mathcal{R}_{\mrm{in},i}^{(\ell)}A_{C/S,i}^{(\ell m)} +
\sum_{\prm{i}=1}^{i-1}\mathcal{R}_{\mrm{in},\prm{i}}^{(\ell)}A_{C/S,\prm{i}}^{(\ell m)}
\end{equation}
and
\begin{equation}
\sum_{\prm{i}=i}^{N_r}\mathcal{R}_{\mrm{out},\prm{i}}^{(\ell)}A_{C/S,\prm{i}}^{(\ell m)} =
\mathcal{R}_{\mrm{out},i}^{(\ell)}A_{C/S,i}^{(\ell m)} +
\sum_{\prm{i}=i+1}^{N_r}\mathcal{R}_{\mrm{out},\prm{i}}^{(\ell)}A_{C/S,\prm{i}}^{(\ell m)}.
\end{equation}

\section{Generalized Algorithm for a multi-patch grid in spherical geometry}
\label{sec:multi-patch-algo}

\begin{figure}
\centering
\tdplotsetmaincoords{60}{140}
\begin{tikzpicture}[tdplot_main_coords,scale=3]

 \pgfmathsetmacro\thtd{-60}
 \pgfmathsetmacro\phid{-140}
 \pgfmathsetmacro\alp{30}
 \pgfmathsetmacro\bet{45}
 \pgfmathsetmacro\gam{30}
 \pgfmathsetmacro\ctht{cos(\thtd)}
 \pgfmathsetmacro\stht{sin(\thtd)}
 \pgfmathsetmacro\cphi{cos(\phid)}
 \pgfmathsetmacro\sphi{sin(\phid)}
 \pgfmathsetmacro\calp{cos(\alp)}
 \pgfmathsetmacro\salp{sin(\alp)}
 \pgfmathsetmacro\calpphi{cos(\alp+\phid)}
 \pgfmathsetmacro\salpphi{sin(\alp+\phid)} 
 \pgfmathsetmacro\cbet{cos(\bet)}
 \pgfmathsetmacro\sbet{sin(\bet)}
 \pgfmathsetmacro\cgam{cos(\gam)}
 \pgfmathsetmacro\sgam{sin(\gam)}
 \pgfmathsetmacro\sclx{1}
 \pgfmathsetmacro\scly{1}
 \pgfmathsetmacro\sclz{1}
 
 \pgfmathsetmacro\sclx{sqrt(( \cphi)^2+(\ctht*\sphi)^2)}
 \pgfmathsetmacro\scly{sqrt((-\sphi)^2+(\ctht*\cphi)^2)}
 \pgfmathsetmacro\sclz{sqrt((-\stht)^2)}
 \coordinate (O)    at (0,0,0);
 \coordinate (X)    at (1/\sclx,0,0);
 \coordinate (Y)    at (0,1/\scly,0);
 \coordinate (Z)    at (0,0,1/\sclz);

 \pgfmathsetmacro\sclx{sqrt(( \calpphi)^2+(\ctht*\salpphi)^2)}
 \pgfmathsetmacro\scly{sqrt((-\salpphi)^2+(\ctht*\calpphi)^2)}
 \pgfmathsetmacro\sclz{sqrt((-\stht)^2)}
 \coordinate (XP)   at ( \calp/\sclx,\salp/\sclx,0);
 \coordinate (YP)   at (-\salp/\scly,\calp/\scly,0);
 \coordinate (ZP)   at (0,0,1/\sclz);

 \pgfmathsetmacro\sclx{sqrt((\cbet*\calpphi)^2+(\cbet*\ctht*\salpphi+\sbet*\stht)^2)}
 \pgfmathsetmacro\scly{sqrt((-\salpphi)^2+(\ctht*\calpphi)^2)}
 \pgfmathsetmacro\sclz{sqrt((\sbet*\calpphi)^2+(\sbet*\ctht*\salpphi-\cbet*\stht)^2)}
 \coordinate (XPP)  at (\cbet*\calp/\sclx,\cbet*\salp/\sclx,-\sbet/\sclx);
 \coordinate (YPP)  at (     -\salp/\scly,      \calp/\scly,           0);
 \coordinate (ZPP)  at (\sbet*\calp/\sclz,\sbet*\salp/\sclz, \cbet/\sclz);
 
 \pgfmathsetmacro\sclx{sqrt((\cgam*\cbet*\calpphi-\sgam*\salpphi)^2
                           +(\cgam*(\cbet*\ctht*\salpphi+\sbet*\stht)+\sgam*\ctht*\calpphi)^2)}
 \pgfmathsetmacro\scly{sqrt((-\sgam*\cbet*\calpphi-\cgam*\salpphi)^2
                           +(-\sgam*(\cbet*\ctht*\salpphi+\sbet*\stht)+\cgam*\ctht*\calpphi)^2)}
 \pgfmathsetmacro\sclz{sqrt((\sbet*\calpphi)^2+(\sbet*\ctht*\salpphi-\cbet*\stht)^2)}
 \coordinate (XPPP) at ( \cgam*\cbet*\calp/\sclx-\sgam*\salp/\sclx, \cgam*\cbet*\salp/\sclx+\sgam*\calp/\sclx,-\cgam*\sbet/\sclx);
 \coordinate (YPPP) at (-\sgam*\cbet*\calp/\scly-\cgam*\salp/\scly,-\sgam*\cbet*\salp/\scly+\cgam*\calp/\scly, \sgam*\sbet/\scly);
 \coordinate (ZPPP) at (                         \sbet*\calp/\sclz,                         \sbet*\salp/\sclz,       \cbet/\sclz);

 \draw[-{Latex[width=3mm]},line width=2.0pt] (O) -- (X) node[anchor=north east]{{\Large $\xi$}};
 \draw[-{Latex[width=3mm]},line width=2.0pt] (O) -- (Y) node[anchor=north west]{\Large $\eta$};
 \draw[-{Latex[width=3mm]},line width=2.0pt] (O) -- (Z) node[anchor=south]{\Large $\zeta$ , $\prm{\zeta}$};

 \draw[-{Latex[width=3mm]},dashed,line width=1.0pt] (O) -- (YP)   node[pos=0.75]{{\AxisRotator[rotate=-9.56,green!50!black]}};

 \draw[-{Latex[width=3mm]},dashed,line width=1.0pt] (O) -- (XP) node[anchor=north east]{{\Large $\prm{\xi}$}};
 \draw[-{Latex[width=3mm]},dashed,line width=1.0pt] (O) -- (YP) node[anchor=west]{{\Large $\prm{\eta}$ , $\pprm{\eta}$}};

 \draw[-{Latex[width=3mm]},dashed,line width=1.0pt] (O) -- (XPP) node[anchor=north]{{\Large $\pprm{\xi}$}};
 
 \draw[-{Latex[width=3mm]},line width=2.0pt,color=blue!40!red!40!white] (O) -- (XPPP) node[anchor=north]{\textcolor{black}{\Large $x$}};
 \draw[-{Latex[width=3mm]},line width=2.0pt,color=blue!40!red!40!white] (O) -- (YPPP) node[anchor=west]{\textcolor{black}{\Large $y$}};
 \draw[-{Latex[width=3mm]},line width=2.0pt,color=blue!40!red!40!white] (O) -- (ZPPP) node[anchor=south]{\textcolor{black}{\Large $z$ , $\pprm{\zeta}$}};

 \draw[-{Latex[width=3mm]},line width=2.0pt] (O) -- (Z)    node[pos=0.75]{{\AxisRotator[rotate=-90,red]}};
 \draw[-{Latex[width=3mm]},line width=2.0pt,color=blue!40!red!40!white] (O) -- (ZPPP) node[pos=0.75]{{\AxisRotator[rotate=-48.59,blue]}};

 \tdplotdefinepoints(0,0,0)(1,0,0)(\calp,\salp,0);
 \tdplotdrawpolytopearc[line width=1.5pt,red]{0.4}{anchor=north east}{\textcolor{red}{\Large $\alpha$}}

 \tdplotdefinepoints(0,0,0)(0,0,1)(\sbet*\calp,\sbet*\salp,\cbet);
 \tdplotdrawpolytopearc[line width=1.5pt,green!50!black]{0.4}{anchor=south east}{\textcolor{green!50!black}{\Large $\beta$}}

 \tdplotdefinepoints(0,0,0)(\cbet*\calp,\cbet*\salp,-\sbet)(\cgam*\cbet*\calp-\sgam*\salp,\cgam*\cbet*\salp+\sgam*\calp,-\cgam*\sbet);
 \tdplotdrawpolytopearc[line width=1.5pt,blue]{0.4}{anchor=north}{\textcolor{blue}{\Large $\gamma$}}

\end{tikzpicture}
\caption{Schematic diagram showing definitions of the Euler angles $\alpha$, $\beta$, and $\gamma$ 
in the $z$-$\prm{y}$-$\pprm{z}$ convention. The Euler angles parametrize the rotational transformation 
from the $[\xi\eta\zeta]$ coordinate system to the $[xyz]$ coordinate system.}
\label{fig:rottrans}
\end{figure}
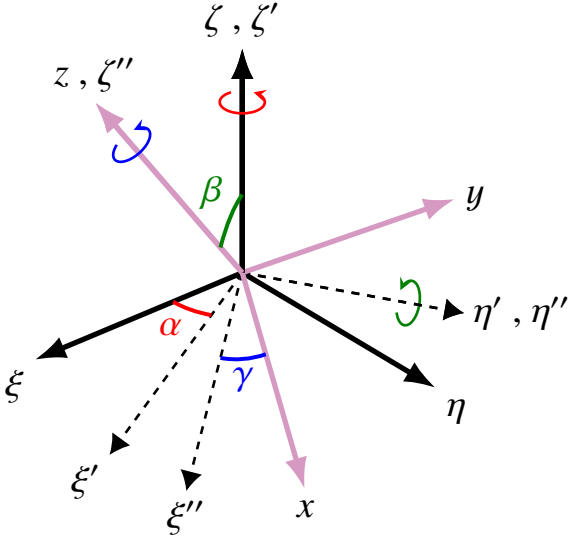

When solving for the gravitational potential on a multi-patch grid configuration in spherical 
geometry, e.g., the Yin-Yang grid \citep{KageyamaSato04} and the cubed-sphere grid 
\citep{Ronchietal96}, the angular integrals over the $4\pi$ spherical surface area in 
Eqs.~(\ref{eq:A_lm}) and (\ref{eq:B_lm}) are splitted into $N_g$ parts with $N_g$ being 
the number of grid patches in the considered grid configuration. Each integration part
is performed over the solid angle $\Omega_g$ covered by the $g^\mrm{th}$ grid patch. 
The functions $\mathcal{A}_{\ell m}$ and $\mathcal{B}_{\ell m}$  can then be re-written as 
\begin{multline}
\mathcal{A}_{\ell m}(r) = \frac{1}{r^{\ell+1}}\sum_{g=1}^{N_g}
\iint_{\Omega_g}\mrm{d}\prm{\Omega}w(\prm{\theta},\prm{\phi})
Y_\ell^{m\,*}(\prm{\theta},\prm{\phi}) \\
\times\int_0^r\mrm{d}\prm{r}(\prm{r})^{\ell+2}\rho(\vect{\prm{r}})
\label{eq:A_lm_mult}
\end{multline}
and
\begin{multline}
\mathcal{B}_{\ell m}(r) = r^{\ell}\sum_{g=1}^{N_g}
\iint_{\Omega_g}\mrm{d}\prm{\Omega}w(\prm{\theta},\prm{\phi})
Y_\ell^{m\,*}(\prm{\theta},\prm{\phi}) \\
\times\int_r^\infty\mrm{d}\prm{r}(\prm{r})^{1-\ell}\rho(\vect{\prm{r}})
\label{eq:B_lm_mult}
\end{multline}
where the complex conjugates of spherical harmonics are multiplied by a weight function 
$w(\theta,\phi)$ to account for overlapping area in the case of an overlapping grid 
configuration. This surface weight function takes a value of $1/n_g(\theta,\phi)$ where 
$n_g(\theta,\phi)$ is the number of grid patches covering an angular position $(\theta,\phi)$.
By definition, $w$ thus equals to 1 in the nonoverlapping region. An integration of this
weight function over the entire surface of all grid patches yields $4\pi$ steradian, i.e.  
\begin{equation}
\sum_{g=1}^{N_g}\iint_{\Omega_{g}}\mrm{d}\prm{\Omega}w(\prm{\theta},\prm{\phi}) = 4\pi.
\end{equation}

Typically, multi-patch grids in spherical geometry are designed such that each patch 
is geometrically identical to ease complications in the implementation of a numerical scheme 
on such grids. Meshes are often constructed using coordinate systems that are local to each 
grid patch. These local coordinate reference frames are related by rotational 
transformation about the common coordinate origin. Evaluation of the angular parts of the 
integrations in Eqs.~(\ref{eq:A_lm_mult}) and (\ref{eq:B_lm_mult}) now become much more 
complicated than in the case of a spherical polar grid since each spherical harmonic mode 
takes a different functional form on each grid patch due to coordinate transformations.
Nevertheless, because under rotational transformation a spherical harmonic of degree $\ell$ 
and order $m$ is simply a linear combination of spherical harmonics of the same degree defined 
in the rotated reference frame and, in addition, because each grid patch is geometrically 
identical the integrations in Eqs.~(\ref{eq:A_lm_mult}) and (\ref{eq:B_lm_mult}) 
can be simplified considerably.

Let $(r,\vartheta_g,\varphi_g)$ be the spherical coordinates of a point $\vect{r}$ in a 
reference frame $[\xi\eta\zeta]^{(g)}$ defined for the construction of the $g^\mrm{th}$ grid 
patch. Since coordinates in reference frames of all grid patches transform only by rotations 
about the coordinate origin the radial coordinate $r$ remains equal for all grid patches, 
and is thus denoted without the subscript $g$. Components of the Cartesian coordinates 
$(\xi_g,\eta_g,\zeta_g)$ are related to the spherical coordinates in the same reference 
frame by the usual coordinate transformation, 
\begin{eqnarray}
\xi_g = r\sin{\vartheta_g}\cos{\varphi_g}, \label{eq:cartosph} \\
\eta_g = r\sin{\vartheta_g}\sin{\varphi_g},
\end{eqnarray}
and
\begin{equation}
\zeta_g = r\cos{\vartheta_g}.
\end{equation}
The corresponding inverse transformation reads 
\begin{eqnarray}
r = \sqrt{\xi_g^2+\eta_g^2+\zeta_g^2}, \\
\vartheta_g = \arccos{(\zeta_g/r)},
\end{eqnarray}
and
\begin{equation}
\varphi_g = \arctan{(\eta_g/\xi_g)}. \label{eq:sphtocar}
\end{equation}

The rotational transformation from the $[\xi\eta\zeta]^{(g)}$ coordinate system to the $[xyz]$ 
coordinate system can be decomposed into three elemental rotations with the amount of 
rotations given by the Euler angles $\alpha_g$, $\beta_g$, and $\gamma_g$. In the
$z$-$\prm{y}$-$\pprm{z}$ convention, the sequence of rotation is 
a rotation by an angle $\alpha_g$ around the $\zeta_g$ axis, then by an angle $\beta_g$ 
around the rotated $\prm{\eta_g}$ axis, and finally by an angle $\gamma_g$ around the 
rotated $\pprm{\zeta_g}$ axis (see Fig.\ref{fig:rottrans}). Correspondingly, the rotational 
transformation matrix relating the Cartesian coordinates $(\xi_g,\eta_g,\zeta_g)$ to 
$(x,y,z)$ is given by 
\begin{align}
\mathfrak{R}^{(g)} & = \mathfrak{R}_{\pprm{\zeta_g}}(\gamma_g)~ 
               \mathfrak{R}_{\prm{\eta_g}}(\beta_g)~
               \mathfrak{R}_{\zeta_g}(\alpha_g) \nonumber \\
& =
\begin{pmatrix} 
 c_{\alpha_g} c_{\beta_g} c_{\gamma_g}-s_{\alpha_g} s_{\gamma_g} & 
 s_{\alpha_g} c_{\beta_g} c_{\gamma_g}+c_{\alpha_g} s_{\gamma_g} & 
-s_{\beta_g}  c_{\gamma_g} \\
-c_{\alpha_g} c_{\beta_g} s_{\gamma_g}-s_{\alpha_g} c_{\gamma_g} & 
-s_{\alpha_g} c_{\beta_g} s_{\gamma_g}+c_{\alpha_g} c_{\gamma_g} & 
 s_{\beta_g}  s_{\gamma_g} \\
 c_{\alpha_g} s_{\beta_g} & 
 s_{\alpha_g} s_{\beta_g} & 
 c_{\beta_g}
\end{pmatrix}
\label{eq:rottransmatrix}
\end{align}
where we abbreviate $\cos{\{\alpha_g,\beta_g,\gamma_g\}}$ and $\sin{\{\alpha_g,\beta_g,\gamma_g\}}$
as $c_{\{\alpha_g,\beta_g,\gamma_g\}}$ and $s_{\{\alpha_g,\beta_g,\gamma_g\}}$, respectively, in the above 
matrix equation for compactness. 

Once the rotation operator is defined relations between spherical harmonics in the $[xyz]$ and 
$[\xi\eta\zeta]^{(g)}$ coordinate systems are given by
\begin{equation}
Y_\ell^m(\theta,\phi) = \sum_{\prm{m}=-\ell}^\ell
D_{\prm{m}m}^\ell(\alpha_g,\beta_g,\gamma_g)Y_\ell^{\prm{m}}(\vartheta_g,\varphi_g)
\label{eq:ylm_transform}
\end{equation}
where coefficients of the linear combinations are elements of the Wigner D-matrix, 
$D_{\prm{m}m}^\ell$ \citep{Wigner31}. Naturally, the Wigner D-matrix is a function of the three 
Euler angles characterizing the rotational transformation between the two coordinate systems. 
With the rotational transformation defined using the $z$-$\prm{y}$-$\pprm{z}$ 
convention, elements of the Wigner D-matrix are expressed as 
\footnote{A number of different notations of the Wigner D-matrix are used in the literature. Here, 
we adopt the same notation as in \citet{MorrisonParker87} where $D_{\prm{m}m}^\ell(\alpha,\beta,\gamma)$
is equal to $D_{\prm{m}m}^\ell(-\alpha,-\beta,-\gamma)$ with the notation that is employed 
in the original work by \citet{Wigner31}.}
\begin{equation}
D_{\prm{m}m}^\ell(\alpha,\beta,\gamma) = e^{-i\prm{m}\alpha}\cdot d_{\prm{m}m}^\ell(\beta)\cdot
e^{-im\gamma}
\label{eq:wigner-D}
\end{equation}
with the reduced Wigner d-matrix, $d_{\prm{m}m}^\ell$ given by
\begin{multline}
d_{\prm{m}m}^\ell(\beta) = \sqrt{(\ell-m)!\,(\ell+m)!\,
(\ell-\prm{m})!\,(\ell+\prm{m})!} \\
\times\sum_{s=0}^{s_\mrm{max}} \left[\frac{(-1)^{\ell-m-s}}
{s!\,(m+\prm{m}+s)!\,(\ell-m-s)!\,(\ell-\prm{m}-s)!}\right. \\ 
\cdot \left.\left(\cos{\frac{\beta}{2}}\right)^{2s+m+\prm{m}}
\left(\sin{\frac{\beta}{2}}\right)^{2\ell-m-\prm{m}-2s}\right].
\label{eq:wigner-d}
\end{multline}
The summation index $s$ is an integer starting from 0 to $s_\mrm{max}$ which is set by  
$s_\mrm{max} = \min{(\ell-m,\ell-\prm{m})}$ such that arguments of the factorials in 
the denominator are always positive.

However, care must be taken when evaluating $d_{\prm{m}m}^\ell(\beta)$ numerically. 
Computation of $d_{\prm{m}m}^\ell(\beta)$ directly using Eq.~(\ref{eq:wigner-d}) 
is known to suffer from serious loss of precision at high degree $\ell$ due to 
cancellation of terms consisting of huge floating point numbers \citep{Tajima15}. 
To circumvent this problem we calculate $d_{\prm{m}m}^\ell(\beta)$ by Fourier decomposition,
which results from factorization of the second elemental rotation of the transformation
\citep{Edmonds64}. The reduced Wigner-d matrix $d_{\prm{m}m}^\ell(\beta)$ of any angle 
$\beta$ can be computed by \citep{TrapaniNavaza06}
\begin{equation}
d_{\prm{m}m}^\ell(\beta)=i^{m-\prm{m}}\sum_{u=-\ell}^\ell d_{u\prm{m}}^\ell(\tpihlf)
d_{um}^\ell(\tpihlf)e^{iu\beta}.
\end{equation}
with the Fourier coefficients computed by utilizing recurrence formulae 
\begin{equation}
d_{\ell 0}^\ell(\tpihlf) = -\sqrt{\frac{2\ell-1}{2\ell}}
d_{(\ell-1)0}^{(\ell-1)}(\tpihlf),
\end{equation}
\begin{equation}
d_{\ell m}^\ell(\tpihlf) = -\sqrt{\frac{\ell(2\ell-1)}{2(\ell+m)(\ell+m-1)}}
\cdot d_{(\ell-1)(m-1)}^{(\ell-1)}(\tpihlf),
\end{equation}
and
\begin{multline}
d_{\prm{m}m}^\ell(\tpihlf) = \frac{2m}{\sqrt{(\ell-\prm{m})(\ell+\prm{m}+1)}}
\cdot d_{(\prm{m}+1)m}^{\ell}(\tpihlf) \\ 
- \sqrt{\frac{(\ell-\prm{m}-1)(\ell+\prm{m}+2)}{(\ell-\prm{m})(\ell+\prm{m}+1)}}
\cdot d_{(\prm{m}+2)m}^{\ell}(\tpihlf).
\end{multline}
The starting condition for these recursions, i.e. the apex of the 
$d_{\prm{m}m}^\ell(\frac{\pi}{2})$ matrix pyramid, is given by 
$d_{00}^0(\frac{\pi}{2})=1$.

We proceed in our derivation by taking the complex conjugate of Eq.~(\ref{eq:ylm_transform}),
and then substituting the result in Eqs.~(\ref{eq:A_lm_mult}) and (\ref{eq:B_lm_mult}). 
The functions $\mathcal{A}_{\ell m}$ and $\mathcal{B}_{\ell m}$ now take the forms  
\begin{multline}
\mathcal{A}_{\ell m}(r) = \frac{1}{r^{\ell+1}}\sum_{g=1}^{N_g}\sum_{\prm{m}=-\ell}^\ell
\left[D_{\prm{m}m}^\ell(\alpha_g,\beta_g,\gamma_g)\right]^* \\ 
\times \iint_{\Omega_g}\mrm{d}\prm{\omega_g}
w(\prm{\vartheta_g},\prm{\varphi_g})Y_\ell^{m\,*}(\prm{\vartheta_g},\prm{\varphi_g})
\int_0^r\mrm{d}\prm{r}(\prm{r})^{\ell+2}\rho(\vect{\prm{r}})
\label{eq:A_lm_trans}
\end{multline}
and
\begin{multline}
\mathcal{B}_{\ell m}(r) = r^{\ell}\sum_{g=1}^{N_g}\sum_{\prm{m}=-\ell}^\ell
\left[D_{\prm{m}m}^\ell(\alpha_g,\beta_g,\gamma_g)\right]^* \\ 
\times \iint_{\Omega_g}\mrm{d}\prm{\omega_g}
w(\prm{\vartheta_g},\prm{\varphi_g})Y_\ell^{m\,*}(\prm{\vartheta_g},\prm{\varphi_g})
\int_r^\infty\mrm{d}\prm{r}(\prm{r})^{1-\ell}\rho(\vect{\prm{r}})
\label{eq:B_lm_trans}
\end{multline}
where $\mrm{d}\omega_g=\sin{\vartheta_g}\,\mrm{d}\vartheta_g\,\mrm{d}\varphi_g$. 
It is important to note that, along with the transformation of spherical harmonics, 
the angular integrals on each grid section in Eqs.~(\ref{eq:A_lm_trans}) and 
(\ref{eq:B_lm_trans}) have also been transformed into integrals in the $[\xi\eta\zeta]^{(g)}$ 
coordinate system that is local to each grid patch. This coordinate transformation will 
allow us to fully exploit the symmetry property of the grid configuration when we 
evaluate these angular integrals numerically.

Inserting these results into Eq.~(\ref{eq:gravpotspherical}), and using identities 
for the reduced Wigner d-matrix \citep{Edmonds64}
\begin{equation}
d_{-\prm{m}-m}^\ell(\beta) = (-1)^{\prm{m}-m}d_{\prm{m}m}^\ell(\beta),
\end{equation}
\begin{equation}
d_{-\prm{m}m}^\ell(\beta) = (-1)^{\ell+m}d_{\prm{m}m}^\ell(\pi-\beta),
\end{equation}
and the identity in Eq.~(\ref{eq:p_l^-m}) for the associated 
Legendre polynomials to eliminate terms containing spherical harmonics with a negative 
order, we obtain, after algebraic rearrangement, an expression for the gravitational 
potential 
\begin{multline}
\Phi(\vect{r}) = -G\sum_{\ell=0}^{\infty}\sum_{m=0}^{\ell}\sqrt{\frac{(\ell-m)!}{(\ell+m)!}}
P_\ell^m(\cos{\theta}) \\
\cdot\left[\cos{(m\phi)}\mathcal{I}_{\ell m}(r)+\sin{(m\phi)}\mathcal{J}_{\ell m}(r)\right].
\label{eq:phimulti}
\end{multline}
The functions $\mathcal{I}_{\ell m}$ and $\mathcal{J}_{\ell m}$ expand into
\begin{equation}
\mathcal{I}_{\ell m}(r) = \mathcal{K}_{CC}^{(\ell m)}(r)+\mathcal{K}_{CS}^{(\ell m)}(r)+
\mathcal{L}_{CC}^{(\ell m)}(r)+\mathcal{L}_{CS}^{(\ell m)}(r)
\label{eq:I_lm}
\end{equation}
and
\begin{equation}
\mathcal{J}_{\ell m}(r) = \mathcal{K}_{SS}^{(\ell m)}(r)-\mathcal{K}_{SC}^{(\ell m)}(r)+
\mathcal{L}_{SS}^{(\ell m)}(r)-\mathcal{L}_{SC}^{(\ell m)}(r)
\label{eq:J_lm}
\end{equation}
with the definitions
\begin{equation}
\mathcal{K_{lm}}^{(\ell m)}(r) = \frac{1}{r^{\ell+1}}\sum_{g=1}^{N_g}\sum_{\prm{m}=0}^\ell
\mathcal{N}_{\mathcal{lm},g}^{(\ell m \prm{m})}\int_0^r\mrm{d}\prm{r}(\prm{r})^{\ell+2}\,
\mathcal{P}_{\mathcal{m},g}^{(\ell\prm{m})}(\prm{r})
\end{equation}
and
\begin{equation}
\mathcal{L_{lm}}^{(\ell m)}(r) = r^{\ell}\sum_{g=1}^{N_g}\sum_{\prm{m}=0}^\ell
\mathcal{N}_{\mathcal{lm},g}^{(\ell m \prm{m})}\int_r^\infty\mrm{d}\prm{r}(\prm{r})^{1-\ell}\,
\mathcal{P}_{\mathcal{m},g}^{(\ell\prm{m})}(\prm{r})
\end{equation}
where each of the symbols $\mathcal{l}$ and $\mathcal{m}$ represent modes $C$ or $S$. 
The integrands $\mathcal{P}_{C,g}^{(\ell\prm{m})}$ and $\mathcal{P}_{S,g}^{(\ell\prm{m})}$ are 
defined by 
\begin{equation}
\mathcal{P}_{C,g}^{(\ell\prm{m})}(r)=\iint_{\Omega_g}\mrm{d}\omega_gw(\vartheta_g,\varphi_g)
P_\ell^{\prm{m}}(\cos{\vartheta_g})\cos{(\prm{m}\varphi_g)}\rho(\vect{r}),
\label{eq:P_C}
\end{equation}
and
\begin{equation}
\mathcal{P}_{S,g}^{(\ell\prm{m})}(r)=\iint_{\Omega_g}\mrm{d}\omega_gw(\vartheta_g,\varphi_g)
P_\ell^{\prm{m}}(\cos{\vartheta_g})\sin{(\prm{m}\varphi_g)}\rho(\vect{r}).
\label{eq:P_S}
\end{equation}
And finally, the four modes normalization factors $\mathcal{N}_{CC,g}^{(\ell m \prm{m})}$, 
$\mathcal{N}_{CS,g}^{(\ell m \prm{m})}$
$\mathcal{N}_{SS,g}^{(\ell m \prm{m})}$, and $\mathcal{N}_{SC,g}^{(\ell m \prm{m})}$ are given by
\begin{multline}
\mathcal{N}_{CC,g}^{(\ell m \prm{m})}=\frac{2}{\lambda_{\prm{m}m}}
\sqrt{\frac{(\ell-\prm{m})!}{(\ell+\prm{m})!}} 
\left\{\cos{(m\gamma_g+\prm{m}\alpha_g)}\cdot d_{\prm{m}m}^\ell(\beta_g)\right. \\
+\mu_{\prm{m}}(-1)^{\ell+m+\prm{m}}\left.\cos{(m\gamma_g-\prm{m}\alpha_g)}
\cdot d_{\prm{m}m}^\ell(\pi-\beta_g)\right\},
\label{eq:N_CC}
\end{multline}
\begin{multline}
\mathcal{N}_{CS,g}^{(\ell m \prm{m})}=\frac{2}{\lambda_{\prm{m}m}}
\sqrt{\frac{(\ell-\prm{m})!}{(\ell+\prm{m})!}} 
\left\{\sin{(m\gamma_g+\prm{m}\alpha_g)}\cdot d_{\prm{m}m}^\ell(\beta_g)\right. \\
-\mu_{\prm{m}}(-1)^{\ell+m+\prm{m}}\left.\sin{(m\gamma_g-\prm{m}\alpha_g)}
\cdot d_{\prm{m}m}^\ell(\pi-\beta_g)\right\},
\label{eq:N_CS}
\end{multline}
\begin{multline}
\mathcal{N}_{SS,g}^{(\ell m \prm{m})}=\frac{2\mu_m}{\lambda_{\prm{m}m}}
\sqrt{\frac{(\ell-\prm{m})!}{(\ell+\prm{m})!}} 
\left\{\cos{(m\gamma_g+\prm{m}\alpha_g)}\cdot d_{\prm{m}m}^\ell(\beta_g)\right. \\
-\mu_{\prm{m}}(-1)^{\ell+m+\prm{m}}\left.\cos{(m\gamma_g-\prm{m}\alpha_g)}
\cdot d_{\prm{m}m}^\ell(\pi-\beta_g)\right\},
\label{eq:N_SS}
\end{multline}
and
\begin{multline}
\mathcal{N}_{SC,g}^{(\ell m \prm{m})}=\frac{2\mu_m}{\lambda_{\prm{m}m}}
\sqrt{\frac{(\ell-\prm{m})!}{(\ell+\prm{m})!}}
\left\{\sin{(m\gamma_g+\prm{m}\alpha_g)}\cdot d_{\prm{m}m}^\ell(\beta_g)\right. \\
+\mu_{\prm{m}}(-1)^{\ell+m+\prm{m}}\left.\sin{(m\gamma_g-\prm{m}\alpha_g)}
\cdot d_{\prm{m}m}^\ell(\pi-\beta_g)\right\}
\label{eq:N_SC}
\end{multline}
with the coefficients 
\begin{equation}
\lambda_{\prm{m}m} = \begin{cases} 2, & \mrm{if}~m=\prm{m}=0, \\
                                1, & \mrm{otherwise} \end{cases}
\end{equation}
and
\begin{equation}
\mu_{m} = \begin{cases} 0, & \mrm{if}~m=0, \\
                       1, & \mrm{otherwise}. \end{cases}
\end{equation}

It is worth noting that the expression for the gravitational potential on a spherical polar 
grid derived in Section~\ref{sec:phi_spherical}
(Eq.~\ref{eq:phispherical}) can easily be recovered by setting
  $N_g=1$ with the Euler angles $\alpha_g=\beta_g=\gamma_g=0$.

\section{Implementation for computation on the Yin-Yang grid}
\label{sec:YYimplementation}

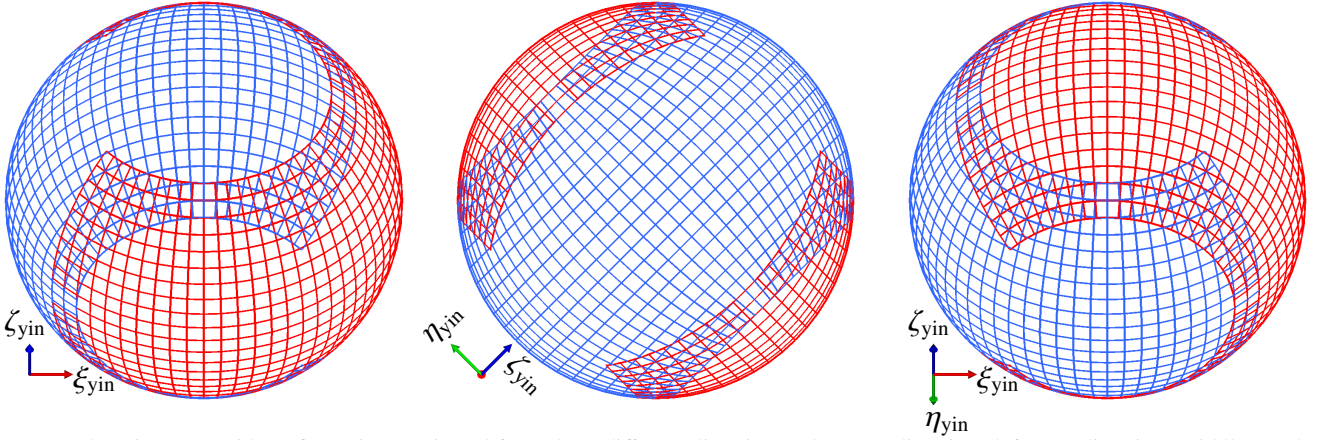
\begin{figure*}
\centering
\begin{overpic}[width=0.32\hsize]{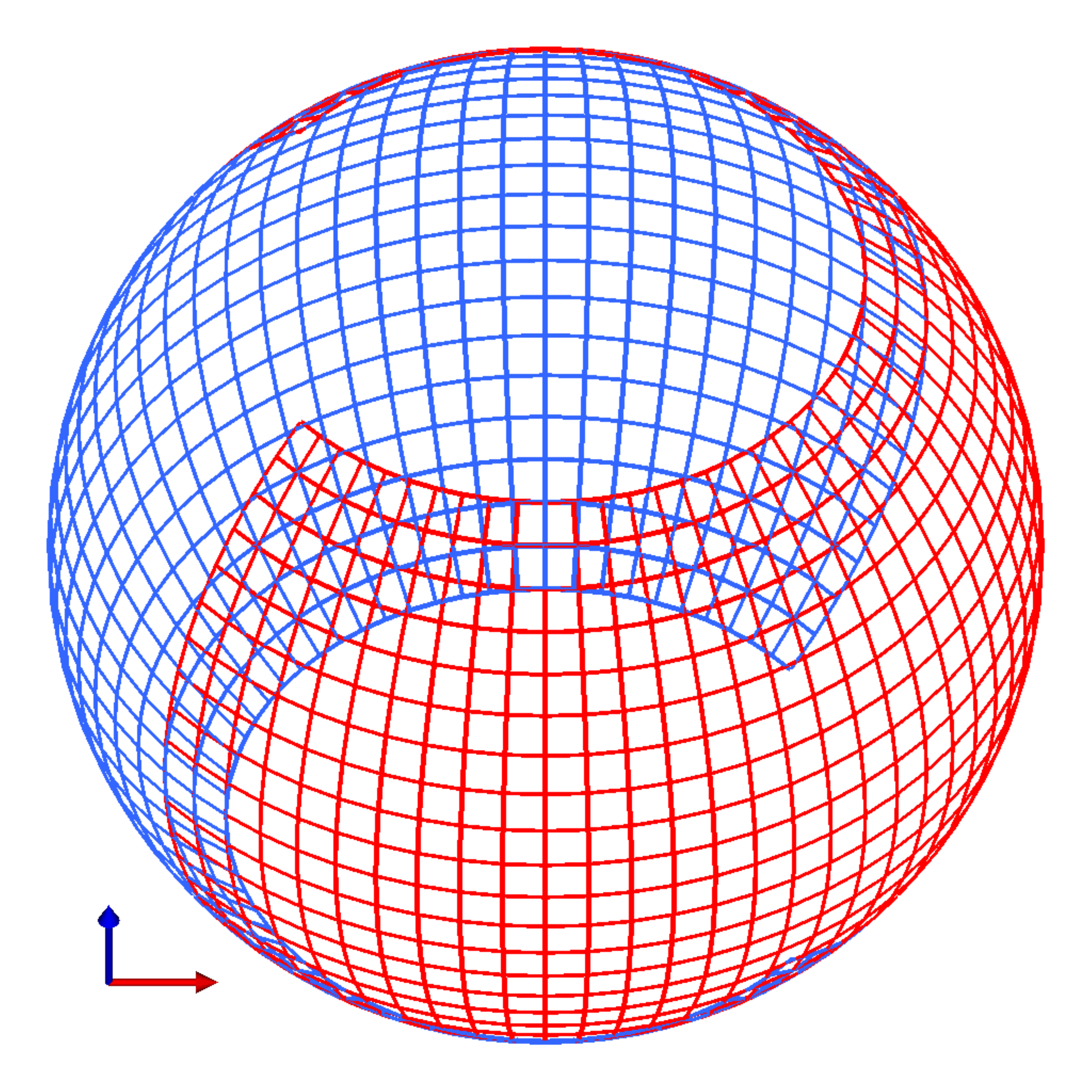}
\put(20,7){\large $\xi_\mrm{yin}$}\put(4,20){\large $\zeta_\mrm{yin}$}
\end{overpic}
\begin{overpic}[width=0.32\hsize]{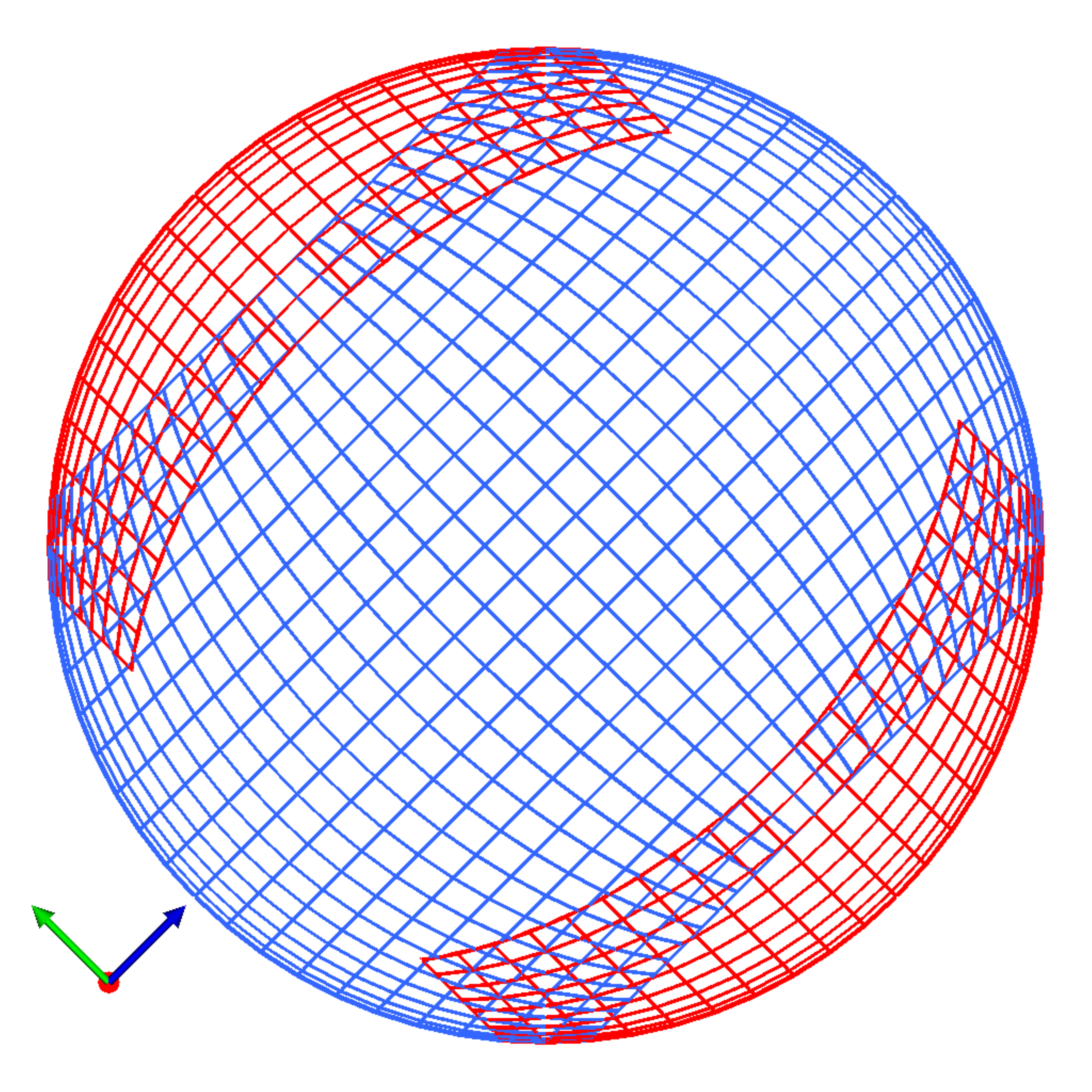}
\put(-4,17){\rotatebox{45}{\large $\eta_\mrm{yin}$}}\put(14,12){\rotatebox{-45}{\large $\zeta_\mrm{yin}$}}
\end{overpic}
\begin{overpic}[width=0.32\hsize]{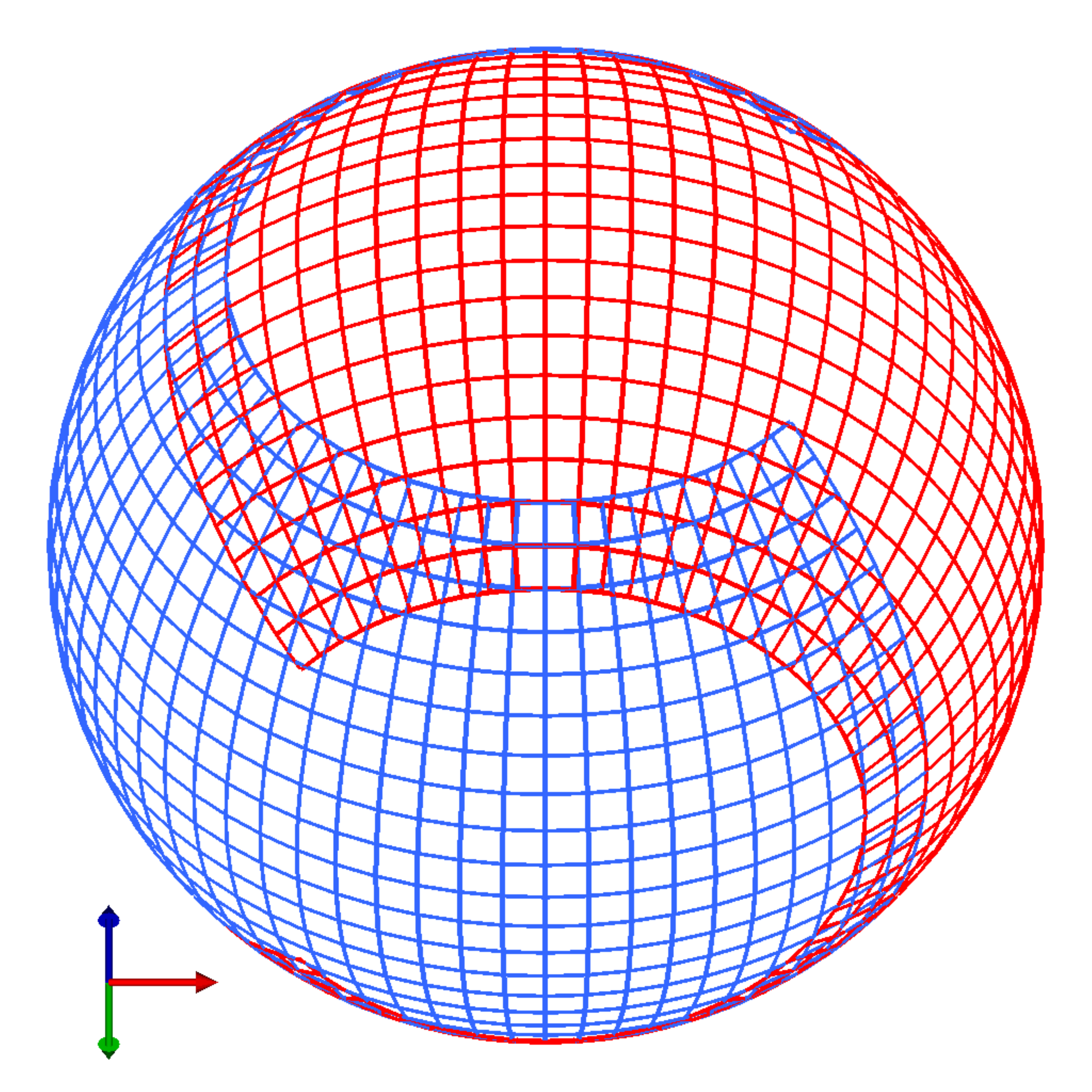}
\put(20,7){\large $\xi_\mrm{yin}$}\put(8,-2){\large $\eta_\mrm{yin}$}\put(4,20){\large $\zeta_\mrm{yin}$}
\end{overpic}
\caption{The Yin-Yang grid configuration as viewed from three different directions: along $+y$-direction 
(left), $+x$-direction (middle), and $+z$-direction (right). The Yin grid is depicted in red, while the Yang 
  grid is shown with blue color. The grid configuration is rotationally symmetric with respect 
to all three viewing axes.}
\label{fig:yygrid}
\end{figure*}

In this section, we demonstrate how the algorithm that we derived in the previous section is 
applied to compute the gravitational potential on the Yin-Yang overset grid configuration. 
The Yin-Yang grid in its most basic configuration consists of two geometrically identical 
overlapping grid patches. Each grid section, Yin or Yang, is simply the low-latitude part 
of the usual spherical polar grid, and therefore forms an orthogonal grid on the surface of 
a sphere. For this particular reason, the algorithm for computation 
of the gravitational potential on the Yin-Yang grid is an easy extension of the base algorithm 
derived for the case of a spherical polar grid in Section~\ref{sec:phi_spherical}.
A pseudo-algorithm providing guidance for implementing our new gravity
solver is presented in Section~\ref{sec:computesteps}.

\subsection{Yin-Yang grid orientation and transformations}

First of all, we construct the Yin and the Yang grid which spans the angular ranges 
\begin{equation}
\frac{\pi}{4}-\Delta \le \vartheta_\mrm{Yin/Yang} \le \frac{3\pi}{4}+\Delta
\end{equation}
and
\begin{equation}
-\frac{3\pi}{4}-\Delta \le \varphi_\mrm{Yin/Yang} \le \frac{3\pi}{4}+\Delta
\end{equation}
in the colatitude and azimuthal directions of their local coordinate reference frames, 
respectively. The angular resolution in both coordinate directions on both grid patches 
is denoted by $\Delta$. We choose a radial grid which is equidistant with a radial grid 
resolution $\Delta_r$. It spans the range 
\begin{equation}
R_\mrm{ib} \le r \le R_\mrm{ob}
\end{equation}
where $R_\mrm{ib}$ and $R_\mrm{ob}$ are the inner and the outer radius of the computational 
domain. Transformation rules between Cartesian and spherical coordinates are given by 
Eqs.~(\ref{eq:cartosph})--(\ref{eq:sphtocar}). Because the Yin-Yang grid configuration is 
symmetric Cartesian coordinates of a point in both the Yin and the Yang coordinate system 
are transformed to coordinates of the other grid patch by a matrix equation of the same 
form, i.e. 
\begin{equation}
\begin{pmatrix} \xi_\mrm{Yin/Yang} \\ \eta_\mrm{Yin/Yang} \\ \zeta_\mrm{Yin/Yang} \end{pmatrix} 
= \begin{pmatrix} -1 & 0 & 0 \\ 0 & 0 & 1 \\ 0 & 1 & 0 \end{pmatrix}
\begin{pmatrix} \xi_\mrm{Yang/Yin} \\ \eta_\mrm{Yang/Yin} \\ \zeta_\mrm{Yang/Yin} \end{pmatrix}. 
\label{eq:caryytransform}
\end{equation}

The Yin-Yang coordinate transformation matrix translates to a rotation about an axis 
$\boldsymbol{\hat{S}}=(\xi_\mrm{Yin/Yang},\eta_\mrm{Yin/Yang},\zeta_\mrm{Yin/Yang})=
(0,\frac{1}{\sqrt{2}},\frac{1}{\sqrt{2}})$ by an angle $\pi$. Hence, instead of aligning the $[xyz]$ 
coordinate reference 
frame with either the reference frame of the Yin or the Yang grid we choose to align the polar axis 
$\boldsymbol{\hat{z}}$ with the axis $\boldsymbol{\hat{S}}$, and define the relative orientation of 
the Yin grid with respect to the $[xyz]$ coordinate reference frame such that 
\begin{equation}
\begin{pmatrix} x \\ y \\ z \end{pmatrix} 
= \begin{pmatrix} 1 & 0 & 0 \\ 0 & \frac{1}{\sqrt{2}} & -\frac{1}{\sqrt{2}} \\ 
0 & \frac{1}{\sqrt{2}} & \frac{1}{\sqrt{2}} \end{pmatrix}
\begin{pmatrix} \xi_\mrm{Yin} \\ \eta_\mrm{Yin} \\ \zeta_\mrm{Yin} \end{pmatrix}. 
\label{eq:yintoxyz}
\end{equation}
Transformation of the Yang coordinates to the $[xyz]$ system can then be obtained by combining 
Eq.~(\ref{eq:caryytransform}) and (\ref{eq:yintoxyz}). This particular choice of orientation 
results in a grid that possesses rotational symmetry of order 2 about all coordinate axes of 
the $[xyz]$ coordinate system (see Fig.~\ref{fig:yygrid}). 

Once the choice of grid orientation for the Yin-Yang grid with respect to the $[xyz]$
coordinate system is defined, the three Euler angles describing the rotational transformation 
from the Yin and the Yang coordinate system to the $[xyz]$ system can be computed by solving 
trigonometric equations resulting from Eqs.~(\ref{eq:rottransmatrix}),(\ref{eq:caryytransform}), 
and (\ref{eq:yintoxyz}). This yields
\begin{equation} 
(\alpha_\mrm{Yin},\beta_\mrm{Yin},\gamma_\mrm{Yin})=(\frac{\pi}{2},\frac{\pi}{4},
\frac{3\pi}{2})
\end{equation}
and
\begin{equation} 
(\alpha_\mrm{Yang},\beta_\mrm{Yang},\gamma_\mrm{Yang})=(\frac{\pi}{2},\frac{\pi}{4},
\frac{\pi}{2}).
\end{equation}
One can see that only the rotation angle $\gamma$ of the third elemental rotation differs 
between the two sets of Euler angles. As a result, this allows us to simplify calculations
of angular weights necessary for computation of the gravitational potential by evaluating 
only one set of weights, either for the Yin or the Yang grid, and obtain the weights for 
the other grid patch by multiplying with a factor 1 or -1. The multiplication factor 
depends on the spherical harmonic mode. This will be demonstrated in the following 
steps.

\subsection{Discretized formulae on the Yin-Yang grid}

As in the case of computation on a spherical polar grid we approximate the density 
distribution inside a grid cell $ijk$ on the Yin and the Yang patch by the cell-averaged 
density $\rho_{ijk,\mrm{Yin/Yang}}$. In addition,we also truncate the summation series over 
spherical harmonics degree at a degree $\ell_\mrm{max}$, and an assumption 
for the weight accounting for overlapping surface area $w$ is applied. When computing 
surface integrals we assume a constant weight within an angular grid zone $jk$, thereby 
replacing the weight function $w$ by a surface-averaged 
value $w_{jk}=1-0.5\alpha_{jk}$ where $\alpha_{jk}$ is the fraction of overlapping surface area 
\citep[see e.g.,][for details]{Wongwathanaratetal10}.

To calculate the gravitational potential at cell vertices of a grid zone $ijk$ on the Yin-Yang 
grid we rewrite the potential in a compact form as
\begin{equation}
\Phi(r_i^+,\vartheta_{j,g}^+,\varphi_{k,g}^+) = 
\sum_{\ell=0}^{\ell_\mrm{max}}\sum_{m=0}^{\ell}
\mathcal{Q}_{C,jk,g}^{(\ell m)}\cdot\mathcal{M}_{C,i}^{(\ell m)}+\mathcal{Q}_{S,jk,g}^{(\ell m)}
\cdot\mathcal{M}_{S,i}^{(\ell m)}.
\label{eq:gravpotmult}
\end{equation}
The gravitational potential expressed in this form reflects directly the actual 
implementation of the method in our numerical code. The prefactors  
$\mathcal{Q}_{\mathcal{l},jk,g}^{(\ell m)}$ are defined by
\begin{equation}
\mathcal{Q}_{C,jk,g}^{(\ell m)} = -G\sqrt{\frac{(\ell-m)!}{(\ell+m)!}}
P_\ell^m(\cos{\theta_{jk,g}^+})\cos{(m\phi_{jk,g}^+)}
\label{eq:Q_C}
\end{equation}
and
\begin{equation}
\mathcal{Q}_{S,jk,g}^{(\ell m)} = -G\sqrt{\frac{(\ell-m)!}{(\ell+m)!}}
P_\ell^m(\cos{\theta_{jk,g}^+})\sin{(m\phi_{jk,g}^+)}
\label{eq:Q_S}
\end{equation}
where $\theta_{jk,g}^+=\theta(\vartheta_{j,g}^+,\varphi_{k,g}^+)$ and 
$\phi_{jk,g}^+=\phi(\vartheta_{j,g}^+,\varphi_{k,g}^+)$, both of which can be computed 
easily by utilizing coordinate transformations. Furthermore, by coordinate transformation 
rules in Eqs.~(\ref{eq:caryytransform}) and (\ref{eq:yintoxyz}), one finds that 
\begin{equation}
\theta(\vartheta_{j,Yang}^+,\varphi_{k,Yang}^+)=\theta(\vartheta_{j,Yin}^+,\varphi_{k,Yin}^+)
\nonumber
\end{equation}
and 
\begin{equation}
\phi(\vartheta_{j,Yang}^+,\varphi_{k,Yang}^+)=
\phi(\vartheta_{j,Yin}^+,\varphi_{k,Yin}^+)+\pi.
\nonumber
\end{equation}
Thus $\mathcal{Q}_{\mathcal{l},jk,\mrm{Yang}}^{(\ell m)}=(-1)^m\mathcal{Q}_{\mathcal{l},jk,\mrm{Yin}}^{(\ell m)}$, thereby 
allowing us to simply store only one set of weights in an actual computation.

On the other hand, the radial weights for reconstruction of the gravitational potential 
at the $n^\mrm{th}$ radial grid interface, $\mathcal{M}_{\mathcal{l},i}^{(\ell m)}$ are 
computed as 
\begin{equation}
\mathcal{M}_{\mathcal{l},i}^{(\ell m)}=\frac{1}{(r_i^+)^{\ell+1}}\sum_{\prm{i}=1}^{i}
\mathcal{R}_{\mrm{in},\prm{i}}^{(\ell)}B_{\mathcal{l},\prm{i}}^{(\ell m)}+(r_i^+)^{\ell}\sum_{\prm{i}=i+1}^{N_r}
\mathcal{R}_{\mrm{out},\prm{i}}^{(\ell)}B_{\mathcal{l},\prm{i}}^{(\ell m)}
\label{eq:M_C}
\end{equation}
with 
\begin{equation}
B_{\mathcal{l},i}^{(\ell m)}=
\sum_{g=\mrm{Yin}}^\mrm{Yang}E_{\mathcal{l},i,g}^{(\ell m)}\equiv
\sum_{g=\mrm{Yin}}^\mrm{Yang}
\sum_{j=1}^{N_\theta}\sum_{k=1}^{N_\phi}\rho_{ijk,g}\mathcal{U}_{\mathcal{l},jk,g}^{(\ell m)}
\label{eq:B_CS}
\end{equation}
and $\mathcal{R}_{\mrm{in},\prm{i}}^{(\ell)}$ and $\mathcal{R}_{\mrm{out},\prm{i}}^{(\ell)}$ defined 
by Eqs.~(\ref{eq:R_in}) and (\ref{eq:R_out}).
The angular weights $\mathcal{U}_{\mathcal{l},jk,g}^{(\ell m)}$ are defined by
\begin{multline}
\mathcal{U}_{C,jk,g}^{(\ell m)}=w_{jk}\sum_{\prm{m}=0}^\ell
\left(\mathcal{N}_{CC,g}^{(\ell m \prm{m})}\mathcal{C}^{(\prm{m})}_k+
\mathcal{N}_{CS,g}^{(\ell m \prm{m})}\mathcal{S}^{(\prm{m})}_k\right)
\mathcal{T}^{(\ell\prm{m})}_j,
\label{eq:U_C}
\end{multline}
and 
\begin{multline}
\mathcal{U}_{S,jk,g}^{(\ell m)}=w_{jk}\sum_{\prm{m}=0}^\ell
\left(\mathcal{N}_{SS,g}^{(\ell m \prm{m})}\mathcal{S}^{(\prm{m})}_k-
\mathcal{N}_{SC,g}^{(\ell m \prm{m})}\mathcal{C}^{(\prm{m})}_k\right)
\mathcal{T}^{(\ell\prm{m})}_j
\label{eq:U_S}
\end{multline}
with $\mathcal{N}_{\mathcal{l}\mathcal{m},g}^{(\ell m \prm{m})}$ defined by 
Eqs.(\ref{eq:N_CC}--\ref{eq:N_SC}).
It is worth noting that the surface weights $w_{jk}$ and the integrals 
$\mathcal{T}^{(\ell\prm{m})}_j$, $\mathcal{C}^{(\prm{m})}_k$, and $\mathcal{S}^{(\prm{m})}_k$ 
(Eqs.\ref{eq:T_lm}--\ref{eq:S_k^m})
take the same values on both the Yin and the Yang grid because of the symmetry of the 
grid configuration and, in addition, because these angular integrals are performed 
using coordinates that are local on each grid patch. 

Finally, it is also important to note that both the prefactors $\mathcal{Q}_{\mathcal{l},jk,g}^{(\ell m)}$ 
and the angular weights $\mathcal{U}_{\mathcal{l},jk,g}^{(\ell m)}$ need to be evaluated only once at 
an initialization step. These coefficients can be re-used to compute the 
gravitational potential of any mass distribution represented on the Yin-Yang grid. 
This is valid under the assumption that the angular grid remains fixed throughout 
the simulation. 

\subsection{Computation steps}
\label{sec:computesteps}

Consider a case in which the Yin-Yang grid is decomposed only in the angular directions
into smaller subdomains. Assume that the number of subdomains equals the number of 
compute tasks, $N_\mrm{tasks}$, being used for computation of the gravitational potential 
on a distributed memory system. For this computational setup the parallelized algorithm 
to compute the gravitational potential using Eq.~(\ref{eq:gravpotmult}) can be summarized 
into the following steps:

\begin{enumerate}
\item Compute and store prefactors $\mathcal{Q}_{\mathcal{l},jk,g}^{(\ell m)}$ 
(Eqs.~\ref{eq:Q_C}--\ref{eq:Q_S}) and 
angular weights $\mathcal{U}_{\mathcal{l},jk,g}^{(\ell m)}$ (Eqs.~\ref{eq:U_C}--\ref{eq:U_S}).
\item Each compute task calculates angular summations 
$E_{\mathcal{l},i,g}^{(\ell m)}=\sum_{j=1}^{N_\theta}\sum_{k=1}^{N_\phi}\rho_{ijk,g}\mathcal{U}_{\mathcal{l},jk,g}^{(\ell m)}$
(Eq.~\ref{eq:B_CS}).
\item Perform a summation of $E_{\mathcal{l},i,g}^{(\ell m)}$ across all compute tasks to 
obtain $B_{\mathcal{l},i}^{(\ell m)}$.
\item Each compute task calculates radial summations $\mathcal{M}_{\mathcal{l},i}^{(\ell m)}$
(Eq.~\ref{eq:M_C})
by using recurrence relations
\begin{equation}
\sum_{\prm{i}=1}^{i}\mathcal{R}_{\mrm{in},\prm{i}}^{(\ell)}B_{\mathcal{l},\prm{i}}^{(\ell m)} =
\mathcal{R}_{\mrm{in},i}^{(\ell)}B_{\mathcal{l},i}^{(\ell m)} +
\sum_{\prm{i}=1}^{i-1}\mathcal{R}_{\mrm{in},\prm{i}}^{(\ell)}B_{\mathcal{l},\prm{i}}^{(\ell m)}
\end{equation}
and
\begin{equation}
\sum_{\prm{i}=i}^{N_r}\mathcal{R}_{\mrm{out},\prm{i}}^{(\ell)}B_{\mathcal{l},\prm{i}}^{(\ell m)} =
\mathcal{R}_{\mrm{out},i}^{(\ell)}B_{\mathcal{l},i}^{(\ell m)} +
\sum_{\prm{i}=i+1}^{N_r}\mathcal{R}_{\mrm{out},\prm{i}}^{(\ell)}B_{\mathcal{l},\prm{i}}^{(\ell m)}.
\end{equation}
\item Multiply $\mathcal{M}_{\mathcal{l},i}^{(\ell m)}$ with prefactors $\mathcal{Q}_{\mathcal{l},jk,g}^{(\ell m)}$ 
and add contributions from all $\ell$ and $m$ moments to reconstruct the gravational 
potential.
\end{enumerate}

To ensure consistency of results when computing the summations 
in step 2 and 3 in parallel we evaluate these summations by using the two-sum 
algorithm \citep{Møller65,Knuth81}. The implementation of the two-sum algorithm 
for the summation between compute tasks (step 3) using the Message Passing 
Interface (MPI) library follows that of \citet{HeDing01}.

The computation steps summarized above are very similar to the paralellized algorithm 
proposed by \citetalias{Almanstoetteretal18}. The differences between the two algorithms 
are as follows: First of all and most importantly, definitions of the angular weights 
$\mathcal{U}_{\mathcal{l},jk,g}^{(\ell m)}$ are different. In \citetalias{Almanstoetteretal18}, the 
angular weights are simply the integrations of spherical harmonics defined 
in the Yin or the Yang coordinate system, which are rotated with respect to each other.
These weights take similar forms as those computed for a spherical polar grid, but
are multiplied by the surface weight factor $w$ to account for grid overlaps. 
Because of this the radial weights $E_{\mathcal{l},i,g}^{(\ell m)}$ for each multipole moment of 
the expansion that results from these angular integrations cannot be directly added 
into one set of radial weights $B_{\mathcal{l},i}^{(\ell m)}$. The algorithm of 
\citetalias{Almanstoetteretal18} thus computes two sets of potential from these radial 
weights in the subsequent steps, and then these are added together in the 
final computation step. That is, the gravitational potential is calculated by considering 
two sources corresponding to the mass density distribution on each grid patch separately. 
On the other hand, the angular weights $\mathcal{U}_{\mathcal{l},jk,g}^{(\ell m)}$ derived in our 
algorithm consider integrations of spherical harmonics defined in a global coordinate 
system that is common for all grid patches. These spherical harmonics functions are  
transformed into linear combinations of spherical harmonics defined in the local 
coordinate system of each grid patch, and are integrated. This transformation is directly 
reflected by the appearance of summations over all spherical harmonics order $\prm{m}$ 
in Eqs.~(\ref{eq:U_C}) and (\ref{eq:U_S}). As a result, our algorithm computes only one set 
of radial weights $B_{\mathcal{l},i}^{(\ell m)}$ which is used by all grid patches in subsequent 
steps to reconstruct the gravitational potential. 

These fundamental differences lead to an improvement of the computational efficiency of 
the new gravity solver. It is easy to see that the operation counts in steps 4 and 5 of 
the algorithm are reduced by a factor of two compared with \citetalias{Almanstoetteretal18}. 
In addition, the size of data communication between MPI processes at step 3 decreases 
by half in comparison to \citetalias{Almanstoetteretal18}. The
algorithm by \citetalias{Almanstoetteretal18} exchanges $2\times
N_\trm{grid}\times N_r \times
\frac{1}{2}[(\ell_\trm{max}+1)^2+\ell_\trm{max}+1]\times
\log_2{N_\trm{tasks}}$ floating-point numbers in total, assuming that
the recursive doubling algorithm \citep{Thakuretal05} is used for the
global reduction operation. In contrast, our new method eliminates the
factor $N_\trm{grid}$ from the expression.

\section{Numerical test and analysis}
\label{sec:results}

\begin{figure}
\centering
\begin{overpic}[width=\hsize]{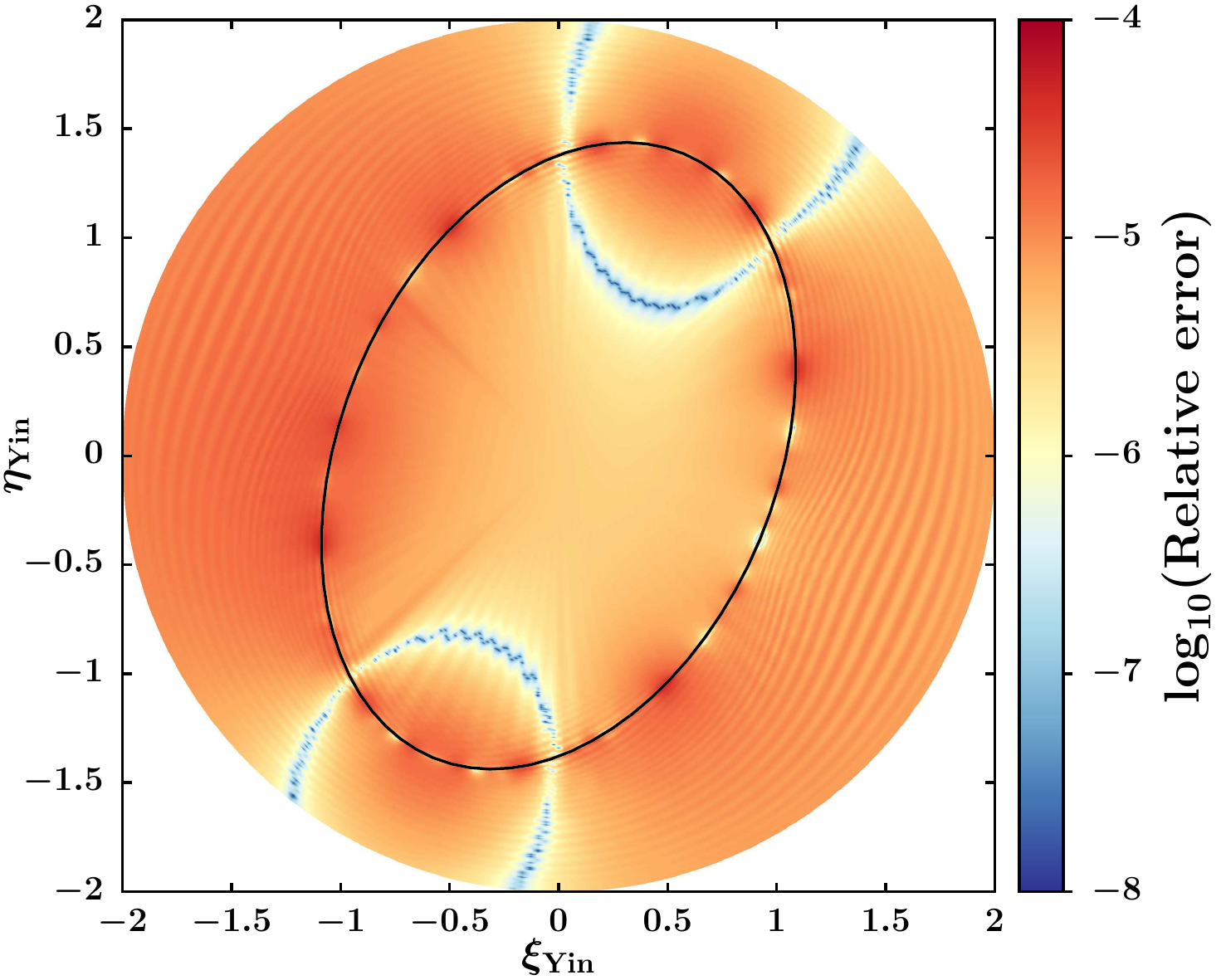}
\end{overpic}
\caption{Pseudocolor plot displaying errors of the gravitational potential of a tri-axial 
ellipsoid of constant density computed on the Yin-Yang grid configuration relative 
to the semi-analytic solution in a cut-plane through the equator of the Yin grid 
section. The angular resolution of the Yin-Yang grid is $1^\circ$ with 800 equidistant 
radial grid zones. The potential is calculated up to a maximum order $\ell_\mrm{max}=80$ 
of the multipole expansion. The black solid line depicts the surface of the ellipsoid. 
The relative errors are shown in logarithmic scale.}
\label{fig:potential}
\end{figure}

\begin{figure}
\centering
\begin{overpic}[width=\hsize]{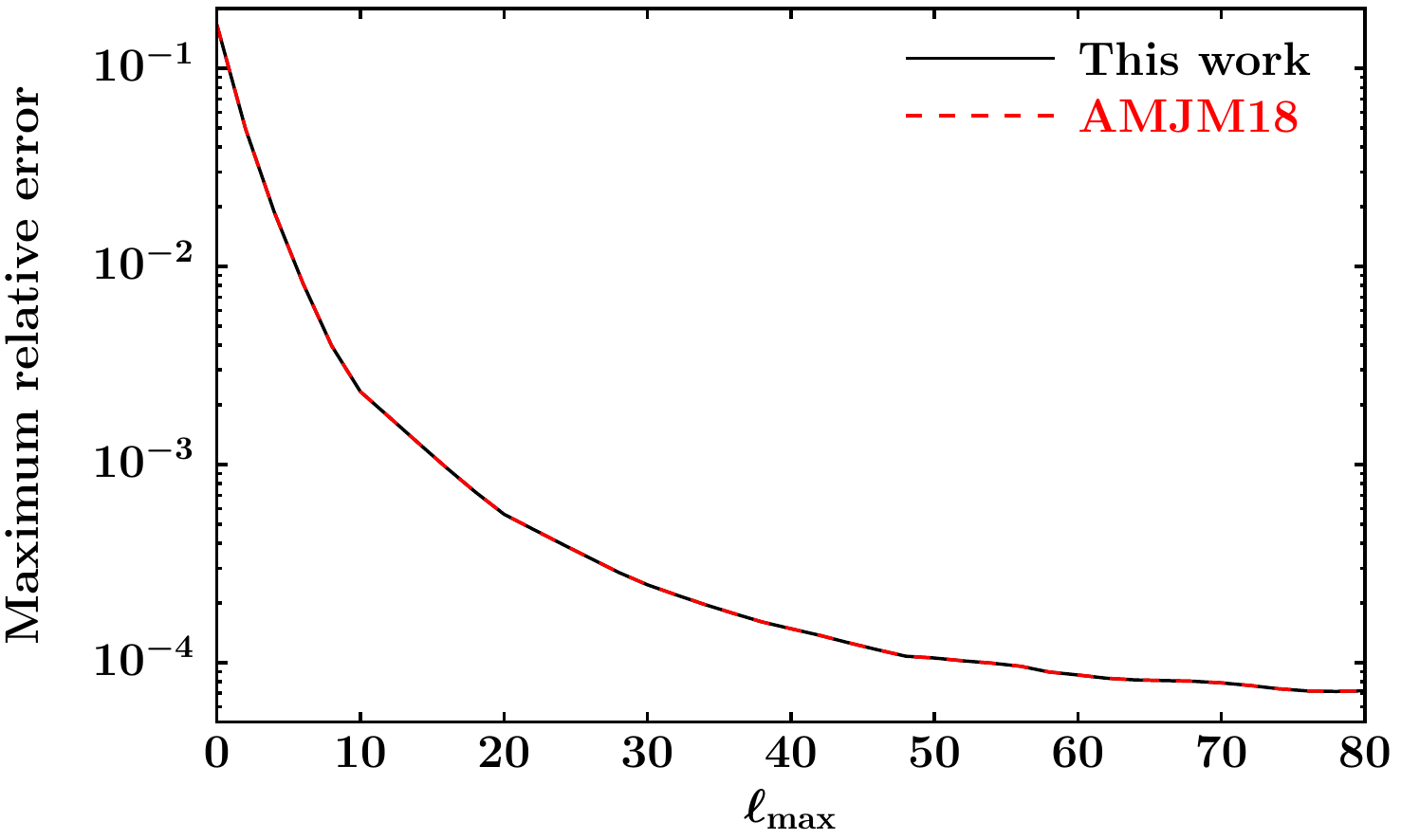}
\end{overpic}
\caption{Maximum error of the gravitational potential of a tri-axial ellipsoid of 
constant density computed on the Yin-Yang grid configuration relative to the semi-analytic 
solution plotted versus the maximum order of the multipole expansion $\ell_\mrm{max}$.
The angular resolution of the Yin-Yang grid is $1^\circ$ with 800 equidistant 
radial grid zones. Results calculated by using our new algorithm (black solid line) 
display excellent agreement with those computed using the algorithm by 
\citetalias{Almanstoetteretal18}
(red dashed line).}
\label{fig:maxerr}
\end{figure}

\subsection{Gravitational potential of a homogeneous ellipsoid}

As a test case for our algorithm we calculate the gravitational potential of 
a homogeneous ellipsoidal body on a Yin-Yang grid setup with an angular resolution $\Delta$ of 
$1^\circ$ and an equidistant radial grid of 800 zones. The radius of the inner and 
outer grid boundary, $R_\mrm{ib}$ and $R_\mrm{ob}$, are set to 0 and 2, respectively. For 
the purpose of comparison with results computed by \citetalias{Almanstoetteretal18} we 
employ the same parameters as listed in their work for this test setup. The 
surface of the tri-axial ellipsoid is defined by the equation
\begin{equation}
\left(\frac{X}{a}\right)^2+\left(\frac{Y}{b}\right)^2+\left(\frac{Z}{c}\right)^2=1
\end{equation}
with parameters for the semi-axes set to $a=1,b=1.5$ and $c=2$. Coordinates $X,Y$ and $Z$ 
relate to Cartesian coordinates in the Yin and the Yang grid system by 
\begin{alignat}{1}
X&=\xi_\mrm{Yin}\cos(\tpieght)-\eta_\mrm{Yin}\sin(\tpieght) \nonumber\\
&=-\xi_\mrm{Yang}\cos(\tpieght)-\zeta_\mrm{Yang}\sin(\tpieght),
\end{alignat}
\begin{alignat}{1}
Y&=\xi_\mrm{Yin}\sin(\tpieght)+\eta_\mrm{Yin}\cos(\tpieght) \nonumber\\
&=-\xi_\mrm{Yang}\sin(\tpieght)+\zeta_\mrm{Yang}\cos(\tpieght),
\end{alignat}
and
\begin{equation}
Z=\zeta_\mrm{Yin}=\eta_\mrm{Yang},
\end{equation}
i.e. the shortest principle axis of the ellipsoid is tilted with respect to the $\xi$-axis 
of the Yin grid by an angle $\frac{\pi}{8}$.
The density $\rho$ for any given point inside the ellipsoidal surface is set to $\rho_0=1$, 
while $\rho=0$ outside of the ellipsoid. 

The analytical solution of the gravitaional potential of this homogeneous ellipsoidal body 
is given by \citep{Chandrasekhar69}. The solution at a point $\vect{R}=(X,Y,Z)$ reads
\begin{equation}
\Phi(\vect{R})=\pi G\rho_0abc\left[\mathfrak{A}(\vect{R})X^2+\mathfrak{B}(\vect{R})Y^2+
\mathfrak{C}(\vect{R})Z^2-\mathfrak{D}(\vect{R})\right]
\label{eq:phiana}
\end{equation}
with the functions $\mathfrak{A}(\vect{R}),\mathfrak{B}(\vect{R}),\mathfrak{C}(\vect{R})$
and $\mathfrak{D}(\vect{R})$ defined by
\begin{equation}
\mathfrak{A}(\vect{R})=\int_{u_0(\vect{R})}^\infty\mrm{d}u\left[(a^2+u)^3(b^2+u)(c^2+u)\right]^{-\frac{1}{2}},
\end{equation}
\begin{equation}
\mathfrak{B}(\vect{R})=\int_{u_0(\vect{R})}^\infty\mrm{d}u\left[(a^2+u)(b^2+u)^3(c^2+u)\right]^{-\frac{1}{2}},
\end{equation}
\begin{equation}
\mathfrak{C}(\vect{R})=\int_{u_0(\vect{R})}^\infty\mrm{d}u\left[(a^2+u)(b^2+u)(c^2+u)^3\right]^{-\frac{1}{2}},
\end{equation}
and
\begin{equation}
\mathfrak{D}(\vect{R})=\int_{u_0(\vect{R})}^\infty\mrm{d}u\left[(a^2+u)^3(b^2+u)(c^2+u)\right]^{-\frac{1}{2}}.
\end{equation}
The value $u_0$ determining the lower limit of the integrations is $u_0=0$
for a point $\vect{R}$ that lies outside of the ellipsoidal surface. On the other hand, in the 
case of a point $\vect{R}$ inside of the ellipsoidal surface, $u_0$ is given by the 
positive root of the equation 
\begin{equation}
\frac{X^2}{a^2+u_0}+\frac{Y^2}{b^2+u_0}+\frac{Z^2}{c^2+u_0}=1.
\label{eq:u0}
\end{equation}
To compute this solution semi-analytically we solve Eq.~(\ref{eq:u0}) for $u_0$ by using 
the bisection method with the tolerance error of $10^{-14}$. Then, the 
functions $\mathfrak{A}(\vect{R}),\mathfrak{B}(\vect{R}),\mathfrak{C}(\vect{R})$
and $\mathfrak{D}(\vect{R})$ are integrated numerically using the Simpson's rule as 
implemented in the {\sc Fortran} subroutine {\sc qsimp} \citep{Pressetal86}. The upper 
limit for the integrations is set to $10^{16}$, and the integrations are evaluated 
up to a fractional accuracy of $5\times10^{-14}$.

Our result for this test is shown in Figure~\ref{fig:potential}, which displays the
color-coded distribution of relative errors in the equatorial slice through the Yin-Yang 
grid. The figure shows errors for the case computed with $\ell_\mrm{max}=80$. The error 
distribution shows a maximum value at the surface of the ellipsoid, which is marked in 
Figure~\ref{fig:potential} by the black solid line to guide the eyes. This is because of 
poor representation of the ellipsoidal surface on the Yin-Yang grid with limited spatial 
resolution. We also compute the maximum relative error as a function of $\ell_\mrm{max}$, 
and show the results in Figure~\ref{fig:maxerr}. The maximum relative error rapidly 
decreases until it approaches an asymptotic value at $\ell_\mrm{max}\sim70$ since the spatial 
discretization error play a more dominant role at high values of $\ell_\mrm{max}$. 
In addition, we implemented the algorithm by \citetalias{Almanstoetteretal18} into our gravity 
solver, and perform the same test with their algorithm. As one can see from the 
curve of the maximum error versus $\ell_\mrm{max}$ in Figure~\ref{fig:maxerr} 
our solver shows excellent agreement with the method by \citetalias{Almanstoetteretal18}
at all values of $\ell_\mrm{max}$. We observe the maximum relative difference between 
results computed with our and their algorithm at a level of $\sim10^{-9}$ only, even at 
high values of $\ell_\mrm{max}$.

\begin{figure}
\centering
\begin{overpic}[width=\hsize]{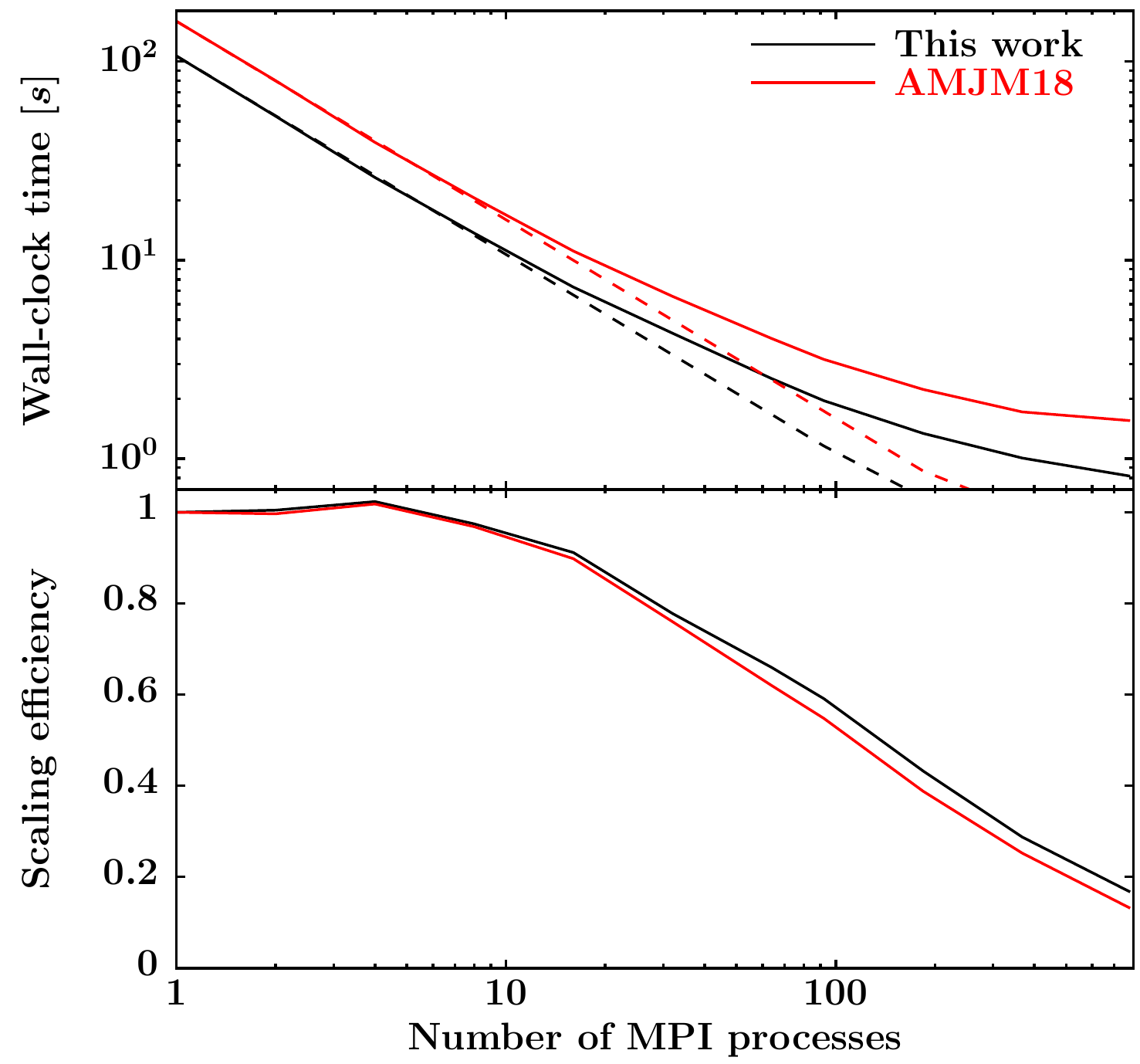}
\end{overpic}
\caption{Averaged wall clock time to solution (top) and strong
  scaling efficiency (bottom) versus the number of MPI processes for
  computation of the gravitational potential on the Yin-Yang grid
  configuration with 800 radial zones and $1^\circ$ angular
  resolution. Black and red solid lines show results calculated with
  our new algorithm and the algorithm by
  \citetalias{Almanstoetteretal18}, respectively. Dashed lines
  represent the ideal scaling behaviour.} 
\label{fig:scaling}
\end{figure}

\begin{table}
\caption{Averaged wall-clock time for computation of the
  gravitational potential on the Yin-Yang grid with 800 radial zones
  and $1^\circ$ angular resolution for different number of MPI
  processes using our algorithm (second column) and the algorithm by 
\citetalias{Almanstoetteretal18} (third column).}
\label{tab:scaling}
\centering
\begin{tabular}{ccc}
\hline
\hline
\multirow{2}{*}{Number of MPI processes} & \multicolumn{2}{c}{Wall-clock time [s]} \\
\cline{2-3}
 & This work & \citetalias{Almanstoetteretal18} \\
\hline
1   & 106.6 & 159.5 \\
2   & 53.07 & 80.02 \\
4   & 26.04 & 39.18 \\
8   & 13.68 & 20.59 \\
16  & 7.310 & 11.10 \\
32  & 4.283 & 6.561 \\
64  & 2.526 & 4.022 \\
92  & 1.960 & 3.164 \\
184 & 1.340 & 2.233 \\
368 & 1.007 & 1.719 \\
782 & 0.816 & 1.553 \\
\hline
\end{tabular}
\end{table}

\subsection{Performance and scaling efficiency}

Table~\ref{tab:scaling} lists the wall-clock time to solution averaged
over 20 calculations of the gravitational potential on the Yin-Yang
grid with $1^\circ$ angular resolution and 800 radial zones using
different numbers of MPI processes. For comparison we show both timing
data for our method and for our implementation of the method by
\citetalias{Almanstoetteretal18}. The data is also plotted in
Fig.~\ref{fig:scaling} (upper panel) along with the strong scaling
efficiency (bottom panel). These numbers are measured using the Intel
Xeon Gold 6148 Processors equipped on the {\sc Cobra} high-performance
computing system at the Max Planck Computing and Data Facility.  

Our data shows that by applying our new algorithm for computation on the Yin-Yang grid 
using a single CPU core the computational efficiency is increased by about 30\% 
when compared with the method by \citetalias{Almanstoetteretal18}. A more detailed analysis 
of both methods reveals that about 70\% of the total computing time is, in fact, spent 
to compute the radial weights $E_{\mathcal{l},i,g}^{(\ell m)}$ in step 2 due to additional costs 
associated with the usage of the two-sum algorithm. Although the operation count at this step is 
equal for both algorithms, we already observe a 30\% gain here. This gain factor results 
from the fact that our algorithm computes only one set of radial weights instead of two 
sets as required by \citetalias{Almanstoetteretal18} method. Consequently, the number of 
load/store instructions in the angular summation loops, which is the computational 
bottleneck in our implementations, is reduced. This demonstrates that it can be misleading 
to compare operation counts when gauging the relative computational efficiency between 
two algorithms. 

As we increase the number of MPI compute tasks we begin to observe benefits from smaller 
data communication volume required by our algorithm when compared with the method by 
\citetalias{Almanstoetteretal18}. While the scaling efficiency of our method is improved 
only slightly relative to the algorithm of \citetalias{Almanstoetteretal18}, the wall-clock 
time to solution is reduced by almost a factor of two when using 782
MPI processes with respect to the method by \citetalias{Almanstoetteretal18}. 
We also point out that although the strong scaling efficiency we report in 
this work is very different from that which is shown by \citetalias{Almanstoetteretal18}, 
cross-comparison of the scaling efficiency should be taken with cautions since it depends 
strongly on details of how an algorithm is implemented and also on details of the 
system running the algorithm. 

\section{Discussions and conclusions}
\label{sec:conclusions}

In this work, we have presented a generalization of the multipole expansion based gravity 
solver by \citet{MuellerSteinmetz95} for efficient computation of the 3D gravitational 
potential on a multi-patch grid configuration in spherical geometry. We derive explicit 
formulae of angular and radial weights for reconstruction of the gravitational potential
by considering integrals of spherical harmonics defined in a global coordinate system 
that is common to all subdomains in the multi-patch grid configuration. These 
spherical harmonics functions are transformed into linear combinations of spherical 
harmonics defined in the local coordinate reference frame of each individual grid patch. 
This transformation eases complications of having to integrate different functions on 
different grid patches when evaluating these angular and radial weights numerically. Linear 
coefficients for the rotational transformation of spherical harmonics 
are given by elements of the well-known Wigner D-matrix \citep{Wigner31} that can be 
evaluated efficiently by recursion relations for any set of Euler angles characterizing 
the transformation between the local and global coordinate system. 

We have applied our new algorithm for calculations of the 3D gravitational potential on 
the Yin-Yang overset grid. Validation of our algorithm is done by comparison of the 
numerical solution to a semi-analytical solution of the gravitational potential of 
a tri-axial ellipsoidal body with homogeneous mass density. For this
test we computed using the maximum degree of the multipole expansion
$\ell_\trm{max}$ of up to 80. At this value of $\ell_\trm{max}$ the
numerical error of the gravitational potential is dominated by the
spatial discretization error associated with the chosen grid
resolution. It is important to note that a suitable choice of
$\ell_\trm{max}$ is problem- and resolution dependent. Judging from
our experiences, in 3D simulations of CCSNe, which is one of the
application areas of our new method, an $\ell_\trm{max}$ of $\sim$20
should already be adequate for typical angular grid resolutions of 1--2 degree.

Our results demonstrate that our algorithm yields a solution that is
as accurate as that obtained by the recent algorithm of
\citetalias{Almanstoetteretal18} proposed for the Yin-Yang
grid. Performance wise, our algorithm benefits from reduced
computational cost and smaller data communiation volume between
parallel compute tasks, thus yielding a faster gravity solver with
better parallel scaling efficiency in comparison with the previous method.

Our new algorithm is easy to implement into an exisiting solver that
is based on the multipole expansion method because it involves only
minor modification to the calculations of angular weights at an
initialization step of the gravity solver. We present detailed
implementation steps of the algorithm for the case of the Yin-Yang
grid configuration in Section~\ref{sec:computesteps}. These
implementation steps can be applied also for computations on other
multi-patch grids in spherical geometry. In the case of a
  non-orthogonal angular grid, computation of angular weights, which
  involves integrations of spherical harmonics, are more
  complicated than the computation on the Yin-Yang grid configuration
  that considers orthogonal angular meshes. Nevertheless, these
  integrations can either be approximated or evaluated by numerical
  integrations. Once the angular weights are computed, the remaining
  steps of the algorithm remain unchanged. 

In a future work, we plan to implement this 
algorithm into our newly developed high-order finite-volume hydrodynamic code, {\sc Apsara}
\citep{Wongwathanaratetal16}, which is capable of dealing with general multi-block 
structured grids in curvilinear coordinates. We also plan to investigate how our algorithm 
can be re-formulated such that it yields higher-order of accuracy of the solution.


\acknowledgments

The author is grateful to Ewald M\"uller for a careful reading of the manuscript, and 
to Ninoy Rahman and Tobias Melson for fruitful discussions. The author
thanks also the anonymous referee for his/her constructive
comments. Computations are carried out on the {\sc Cobra} 
high-performance computing system at the Max Planck Computing and Data Facility.

\software{VisIt \citep{HPV:VisIt}}

\bibliographystyle{aasjournal}
\bibliography{gravity_solver}

\end{document}